\documentclass[a4paper,fleqn]{cas-dc}

\usepackage[numbers]{natbib}
\usepackage{stfloats}
\usepackage{subfigure}
\usepackage{graphicx}
\usepackage{diagbox}
\usepackage{float}
\usepackage{soul}
\usepackage{color}

\def\tsc#1{\csdef{#1}{\textsc{\lowercase{#1}}\xspace}}
\tsc{WGM}
\tsc{QE}
\tsc{EP}
\tsc{PMS}
\tsc{BEC}
\tsc{DE}

\begin{document}

\let\WriteBookmarks\relax
\def\floatpagepagefraction{1}
\def\textpagefraction{.001}

\newcommand{\ravi}[1]{{\sethlcolor{green}}{\hl{#1}}}

\title [mode = title]{CN-Celeb: multi-genre speaker recognition}
\tnotemark[1]
\tnotetext[1]{This document was supported by the National Natural Science Foundation of
China (NSFC) under Grants No.61633013 and No.62171250. Dong Wang (wangdong99@mails.tsinghua.edu.cn) is the corresponding author.
Lantian Li, Ruiqi Liu and Jiawen Kang are joint first authors.}
\author[1]{Lantian Li}
\author[1,2]{Ruiqi Liu}
\author[1,3]{Jiawen Kang}
\author[6]{Yue Fan}
\author[6]{Hao Cui}
\author[4]{Yunqi Cai}
\author[5]{Ravichander Vipperla}
\author[1]{Thomas Fang Zheng}
\author[1]{Dong Wang}
\address[1]{Center for Speech and Language Technologies (CSLT), BNRist at Tsinghua University, Beijing}
\address[2]{China University of Mining and Technology-Beijing}
\address[3]{Department of Systems Engineering and Engineering Management, Chinese University of Hong Kong}
\address[4]{Department of Computer Science and Technology, Tsinghua University}
\address[5]{Samsung AI Center, Cambridge, UK}
\address[6]{Key Laboratory of Transient Physics, Nanjing University of Science and Technology}

\begin{abstract}
Research on speaker recognition is extending to address the vulnerability in the wild conditions,
among which genre mismatch is perhaps the most challenging,
for instance, enrollment with reading speech while testing with conversational or singing audio.
This mismatch leads to complex and composite inter-session variations,
both intrinsic (i.e., speaking style, physiological status) and
extrinsic (i.e., recording device, background noise).
Unfortunately, the few existing multi-genre corpora are not only limited in size
but are also recorded under controlled conditions,
which cannot support conclusive research on the multi-genre problem.
In this work, we firstly publish CN-Celeb, a large-scale multi-genre corpus
that includes in-the-wild speech utterances of 3,000 speakers in 11 different genres.
Secondly, using this dataset, we conduct a comprehensive study on the
multi-genre phenomenon, in particular the impact of the multi-genre challenge on speaker recognition
and the performance gain when the new dataset is used to conduct multi-genre training.
\end{abstract}


\maketitle

\section{Introduction}

Speaker recognition aims to verify the claimed identity of a person using her/his spoken utterance as input modality.
With several decades of research, the performance of speaker recognition systems has been remarkably improved, and
commercial usage has been made feasible in certain conditions ~\cite{campbell1997speaker,reynolds2002overview,hansen2015speaker}.

A long-standing theme in speaker recognition research has been the problem of tackling various speaker-independent variations in the speech signal.
These variations could be extrinsic or intrinsic. The most significant extrinsic variations
include diversity in recording device, ambient acoustics, background noise, transmission channel, and distortions introduced in pre-processing algorithms.
Intrinsic variations refer to both the minor and universal randomness in the movement of pronunciation apparatus as well as
more explicit diversity in speaking style (e.g., reading or spontaneous), speaking rate, emotion, and physical status.
These variations pose the major challenge for speaker recognition systems.

The history of speaker recognition research can be seen as a pursuit towards solving the impact of these variations on the recognition accuracy.
For instance, initial research was constrained to text-dependent tasks and focused on solving the variation caused by pronunciation randomness, in which the Hidden Markov Model (HMM) was the most popular~\cite{parthasarathy1996general}.
Later research attempted to solve text-independent tasks and had to deal with phonetic variation, which
boomed the Gaussian Mixture Model with Universal Background Model (GMM-UBM) architecture~\cite{Reynolds00}.
Further research tried to address inter-session variation caused by channels and speaking styles, for which the i-vector/PLDA architecture was the most successful~\cite{dehak2011front}.
Recently, the research focus has been targeted towards dealing with complex variations in the \emph{wild} scenarios,
for which deep learning methods have been demonstrated to be the most powerful~\cite{ehsan14,li2017deep,snyder2018xvector,okabe2018attentive}.

Multi-genre scenario is perhaps the most challenging scenario for speaker recognition, as it involves nearly
all the complex variations one can imagine. For example, a speaker aged 20 may be registered with the system using an interview speech, recorded in a quiet environment, with a relatively formal speaking style, and using a far-field table microphone; while the test may be with a singing speech of the same speaker at age 40, in a live show under music background, with a close-talk hand-held microphone.
From another perspective, multi-genre is not an artificial challenge,
it is indeed encountered in many real-life applications.
For instance, a good speaker recognition system must be able to accept a user several months after the registration,
even if the user uses a different cellphone and speaks in a different style. In summary, we argue that
good performance on multi-genre scenarios is a sufficient and necessary condition
for practical success of speaker recognition research.

Unfortunately, the existing state-of-the-art techniques perform poorly in multi-genre conditions.
Our recent experimental results show that
a system trained with the most popular recipe and using the largest corpus publicly available (VoxCeleb)
performed quite poorly on CN-Celeb1~\cite{fan2020cn}, a multi-genre corpus that we have recently published.
Specifically, the results in terms of Equal Error Rate (EER) was 3.75\% on SITW, the accompanying test data of VoxCeleb.
On CN-Celeb.E, a subset of CN-Celeb1 consisting of 200 speakers,
the EER result was 15.52\%, an increase of 300\% compared to the SITW result.

The importance of multi-genre scenario and the below-par performance of existing techniques in this scenario necessitates a concentrated research effort in this direction. However, the present CN-Celeb1 corpus~\cite{fan2020cn} is insufficient to derive comprehensive  research conclusions due to its limited data size.
CN-Celeb1 has only 1,000 speakers and 270 hours of speech signals
in total. The limited data size makes it hard to be used as a standalone training set, especially when the models are based on deep learning methods. This has been clearly demonstrated in our previous experiments~\cite{fan2020cn} where the performance of i-vector system was better than that of x-vector system (14.24\% vs. 14.78\%),
when the two systems were trained with 800 speakers (CN-Celeb1.T)
and tested on the rest 200 speakers (CN-Celeb.E).
Moreover, the performance of x-vector system trained with CN-Celeb1.T was even worse than the one trained with VoxCeleb,
even though the former is based on multi-genre data and so is under matched condition.
The data insufficiency is an obstacle for researchers for a deep investigation into the multi-genre challenge.

In this work, we firstly publish a new large-scale multi-genre corpus, called CN-Celeb2.
CN-Celeb2 shares the same 11 genres as CN-Celeb1, but the data size is much larger.
It contains over 520,000 utterances from 2,000 Chinese celebrities.
The two multi-genre corpora make up the overall CN-Celeb corpus\footnote{The dataset can be downloaded from
https://openslr.org/82/}, which can be used to perform fully multi-genre training and test.
Secondly, based on the new CN-Celeb database, we conduct a comprehensive study
on multi-genre speaker recognition. In particular, we employ multi-genre training to improve
model robustness in cross-genre scenarios, and also investigate the efficacy of a meta-learning approach
to improve model generalization for novel genres.

\section{Speaker recognition: challenge, technique and data}

We firstly present a historical review of speaker recognition research.
Different from the previous overviews that concentrate on details of speaker recognition techniques~\cite{campbell1997speaker,reynolds2002overview,hansen2015speaker},
this review focuses on the interaction amongst scientific challenge, technical development and data accumulation.
With this outlook, we categorize the development of speaker recognition research into 4 phases, as outlined in Table~\ref{tab:tech}.

\begin{table*}[bp]
  \caption{History of speaker recognition research.}
   \label{tab:tech}
   \centering
   \begin{tabular}{lllll}
     \cmidrule{1-5}
         Phases & Approximate Period       &  Variations      &  Techniques             & Typical Corpora             \\
     \cmidrule{1-5}
         I      & 1970-2000  &  pronunciation   &  DTW, HMM               & Private data                \\
         II     & 1995-2010  &  phone           &  GMM-UBM                & Switchboard, NIST SRE (-03) \\
         III    & 2005-2016  &  session         &  JFA, i-vector/PLDA     & Mixer, NIST SRE (04-16)     \\
         IV     & 2017-      &  complex         &  DNN                    & VoxCeleb, NIST SRE (18-19)  \\
     \cmidrule{1-5}
   \end{tabular}
\end{table*}

\subsection{Phase 1: Randomness in pronunciation}

The initial foray into speaker recognition focused on the randomness in pronunciation.
One cannot produce the same words/utterances in exactly the same way.
Early speaker recognition research focused on solving this type of variation, by using either non-parametric methods
such as Dynamic Time Warping (DTW)~\cite{rosenberg1976automatic} or with parametric models such as Hidden Markov Model (HMM)~\cite{parthasarathy1996general,matsui1993concatenated}.

Researchers in this period often used small self-collected datasets. For example,
Doddington recorded 123 male speakers in a sound booth using a dynamic microphone, where 63 males
were used for target trials and the rest 60 males for imposter trials~\cite{rosenberg1976automatic}.
Similarly, Parthasarathy et al. recorded 51 males and 49 females over long distance telephony channel, and each speaker made 26 calls uttering the same phrase~\cite{parthasarathy1996general}.

\subsection{Phase 2: Phonetic variation}

Further investigations attempted to deal with the phonetic variation -- the main obstacle towards text-independent
speaker recognition. GMM~\cite{reynolds1995automatic} and its adapted version, GMM-UBM~\cite{Reynolds00}
were demonstrated to be the most effective towards this end.

A large dataset is necessary to train the Gaussian components, and so data requirement during this phase was more demanding than Phase 1.
Moreover, the data used by researchers began to be more standardized, partly due to the NIST
Speaker Recognition Evaluation (SRE) started in 1996~\cite{alvin2004nist}.
One of the most popular corpora during this phase was Switchboard collected by the Linguistic Data Consortium (LDC)\footnote{https://www.ldc.upenn.edu/}.
This corpus incorporated several collections, each of which includes hundreds of speakers and thousands of conversations.
It was extensively used in the NIST SRE series from 1996-2003,
and also formed an important part of the training set in the later NIST SREs.

\subsection{Phase 3: Session variation}

Session variation refers to the systematic change in speaking style or acoustic condition when the same speaker speaks
in different sessions.
Kenny's important work on Joint Factor Analysis (JFA)~\cite{kenny2003new,Kenny07,kenny2005joint} paved the way for solving
general variations between sessions.
The i-vector model, a successor of JFA, made a further leap~\cite{dehak2011front} in producing session-based vectors that involve all types of long-term variations, and leave the task of discriminating different types of variations to a back-end model.
Numerous results have demonstrated that the i-vector model could achieve very good performance with
accompanying probabilistic linear discriminant analysis (PLDA)~\cite{Ioffe06,prince2007probabilistic} as its back-end model.

The data requirement to solve session variation is much more demanding than the previous two phases.
Especially, differentiating session variation requires single-speaker multiple-condition (SSMC) data.
This type of data is much harder to collect as it requires the same speaker providing utterances under different conditions.
Fortunately, NIST SRE, after year by year evaluation, offered a large amount of SSMC data, and the
research conducted in this phase mostly used the NIST SRE data, which was primarily collected by LDC
under the Mixer protocol~\cite{cieri2007resources}.

\subsection{Phase 4: Complex variation}

The session variation, especially covered by the NIST SRE, varies in channels, background noises and even languages.
However, these variations are largely under control. The speech data collected by the Mixer project,
be it telephonic conversations or interviews, was constrained by the collection process, e.g.,
the participants were fully cooperative.

Recently, researchers are attempting to solve a more challenging task: recognizing speakers \emph{in the wild}.
A key feature of this task is that the speakers do not cooperate with or are even aware of being recorded,
and the recording conditions are fully unconstrained, leading to more complex variation.
For example, the audio/video posted on YouTube may be fully spontaneous and recorded by diverse devices.

Addressing such complex variations is highly challenging. Fortunately, the DNN-based methods have shown great potential
in dealing with this problem~\cite{ehsan14,li2017deep,snyder2018xvector,okabe2018attentive}.
So far, the most popular deep learning architecture is based on the concept `deep embedding', which converts
speech segments of variable lengths to fixed-length continuous vectors. Accompanied by a back-end scoring model (e.g., PLDA),
the deep embedding approach has gained the state-of-the-art performance.

The most successful deep embedding model is the x-vector model, proposed by Snyder et al.~\cite{snyder2018xvector}.
Recent progress on the deep speaker embedding approach includes
more comprehensive architectures~\cite{chung2018voxceleb2,Jung2019raw},
improved pooling methods~\cite{okabe2018attentive,Cai2018,Chen2019tied,xie2019utterance},
better training criteria~\cite{li2016max,ding2018mtgan,Wang2019centroid,bai2019partial,Gao2019improving,Zhou2019deep},
and better training schemes~\cite{Li2019boundary,Wang2019phonetic,Stafylakis2019}.
Another popular deep learning approach is the end-to-end modeling, which discriminates speakers of two speech segments
directly~\cite{heigold2016end,zhang2016end,rahman2018attention}. A key advantage of the end-to-end approach is that the
training and test are based on the same criterion, which ensures the test is optimal if the data is sufficient and the
training can be well conducted. However, the training process is often tricky~\cite{wang2017need}.

Due to the data-driven nature, DNN-based methods are data-hungry. Ideally, the training data must comprise
all the potential variations and their combinations that may appear in real applications.
To meet this demand, researchers from SRI released SITW,
the first dataset in unconstrained conditions~\cite{mclaren2016speakers}.
This dataset contains audio from 299 speakers, with an average of 8 different sessions per speaker.
Although very valuable, SITW is too small to be used as a training set.
Oxford released a large in-the-wild corpus VoxCeleb1 in 2017~\cite{nagrani2017voxceleb},
and an even larger one VoxCeleb2 in 2018~\cite{chung2018voxceleb2}.
The total number of speakers in the two corpora exceeds 7,000,
which is fairly large for speaker recognition research.
More importantly, both SITW and VoxCeleb are free, which significantly promoted the recent research on
complex and unconstrained conditions.

\subsection{Data is still insufficient}

With decades of research, complex variation can be partially addressed, thanks to the SITW and VoxCeleb corpora.
However, the present research is yet to solve the truly complex (and difficult) variations. In fact,
most of SITW/VoxCeleb data are from interviews, so the variation on speaking styles has been largely constrained.
Most importantly, the recording condition and speaking style
of each speaker in these corpora do not change much, which means that speakers and
conditions may be heavily coupled~\cite{shonmultimodal}.

As we have mentioned, the multi-genre scenario involves the truly complex variation, making it one of the most difficult
conditions for speaker recognition research\footnote{Others may include recognition with disguised speech or non-speech signals
such as laugh and cough~\cite{zhang2018human}.}. Unfortunately, none of existing datasets is
really multi-genre, including SITW and VoxCeleb, which makes the research on multi-genre speaker recognition
nearly impossible.
The recently published CN-Celeb1 corpus, allows some preliminary studies in this direction~\cite{kang2020domain,kataria2020analysis,mo2020weighted,chen2021self}.
However, the lack of multi-genre \emph{training} data severely precludes further study on this important subject.
We therefore start by building a large-scale multi-genre training set, and then study some simple
techniques to address the multi-genre challenge.

\section{CN-Celeb2: features and collection pipeline}

\subsection{Revisit CN-Celeb1}

The CN-Celeb1 corpus is a free and public speaker recognition dataset released by the research group of the authors~\cite{fan2020cn}.
The speech data were collected from Bilibili, a public media source, using an automated pipeline similar to the one used to collect
VoxCeleb~\cite{nagrani2017voxceleb}. Human check was arranged to ensure the quality of the collected data.
Especially, CN-Celeb1 was intentionally designed to cover multiple genres, in particular cross-genre situations.
The entire dataset contains more than 130,000 utterances from 1,000 Chinese celebrities, and covers 11 different genres in real world.
Readers can refer to the original publication~\cite{fan2020cn} for more details of the data profile and the collection pipeline.

We note that choosing celebrities as the target speakers is important for CN-Celeb1 to achieve its goal.
Celebrities naturally appear in multiple situations and speak in multiple genres, which
perfectly matches our research on multi-genre phenomenon.
However, we do not expect that the speaking style of celebrities would be fully spontaneous,
and we can neither guarantee that speech of celebrities can perfectly represent that of the general public.
A 100\% spontaneous speech dataset covering a large population and diverse genres is certainly valuable,
but constructing such a dataset will be very difficult, considering the constraints in terms of legislation and technical possibility.
At present, although not fully spontaneous and representative, speech of celebrities is sufficient for our research purpose.

Recently, CN-Celeb1 has attracted increasing attention and several multi-genre studies have been carried out using this
dataset~\cite{kang2020domain,kataria2020analysis,mo2020weighted,chen2021self}.
Although very valuable, CN-Celeb1 is not large enough to be used as a standalone training set.
We therefore present a new large-scale multi-genre corpus called CN-Celeb2 to meet the requirements for multi-genre training.
This section summarizes the data collection pipeline and presents the data profile of this new corpus.
CN-Celeb1 and CN-Celeb2 make up the overall CN-Celeb corpus. It has been published in OpenSLR\footnote{https://openslr.org/82/} and is freely available for researchers.

\subsection{Data Description}

CN-Celeb2 shares the same features as CN-Celeb1. Both the corpora were collected from Chinese open media,
and all the constituent speakers are Chinese celebrities. The overall statistics are shown in Table~\ref{tab:comp}.

\begin{table}[htbp]
  \centering
  \caption{Comparison between \emph{CN-Celeb1} and \emph{CN-Celeb2}.}
  \label{tab:comp}
  \begin{tabular}{lll}
   \cmidrule{1-3}
                    & CN-Celeb1     & CN-Celeb2  \\
   \cmidrule{1-3}
    Language        & Chinese       & Chinese    \\
    Genre           & 11            & 11         \\
    \# of Sources    & 1             & 5          \\
    \# of Spks      & 1,000         & 2,000      \\
    \# of Utters    & 130,109       & 529,485    \\
    \# of Hours     & 274           & 1,090      \\
    \# of SSMC Spks & 745           & 658        \\
    Human Check     & Yes           & Yes        \\
   \cmidrule{1-3}
  \end{tabular}
\end{table}

\begin{itemize}
\item The data volume of CN-Celeb2 is larger than CN-Celeb1.
CN-Celeb2 contains $529,485$ utterances from $2,000$ Chinese celebrities, the total speech duration is $1,090$ hours,
which is around 4 times the volume of CN-Celeb1.

\item CN-Celeb2 was collected from more media sources compared to CN-Celeb1.
All the data of CN-Celeb1 were collected from Bilibili\footnote{https://bilibili.com}. For CN-Celeb2,
we collected singing data from NetEase Cloud\footnote{https://music.163.com/} and Changba\footnote{https://changba.com/},
recitation data from Himalaya\footnote{https://www.ximalaya.com/}, and vlog data from Tik Tok\footnote{https://www.douyin.com/}.
Fig.~\ref{fig:source} shows the source distribution of the two datasets.

\begin{figure}[htbp]
\centering
\includegraphics[width=1\linewidth]{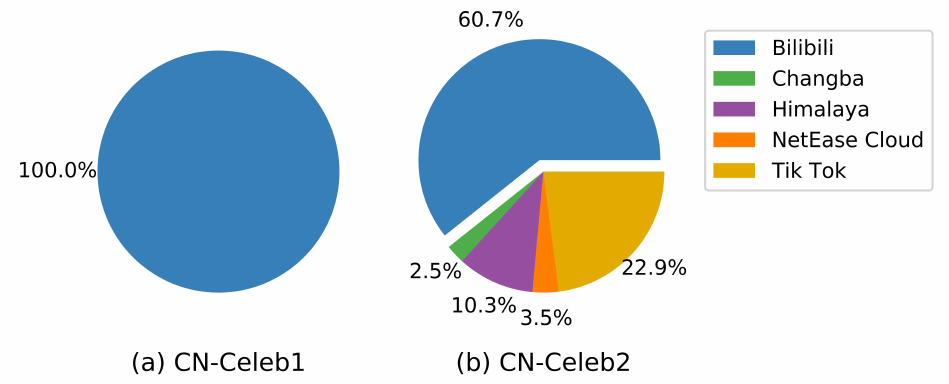}
\caption{The media source distribution of (a) \emph{CN-Celeb1} and (b) \emph{CN-Celeb2}.}
\label{fig:source}
\end{figure}

\item The audio duration distributions of CN-Celeb1 and CN-Celeb2 are shown in Table~\ref{tab:length}.
It can be seen that short utterances account for a larger proportion in both CN-Celeb1 and CN-Celeb2,
which reflects the scenario of most real-life applications but leads to a more complex challenge for speaker recognition research.

\begin{table}[htp]
  \caption{The audio duration distribution of \emph{CN-Celeb1} and \emph{CN-Celeb2}.}
  \label{tab:length}
  \scalebox{0.88}{
  \begin{tabular}{lllll}
    \cmidrule{1-5}
  \multirow{2}{*}{Duration(s)}  &  \multicolumn{2}{c}{CN-Celeb1} & \multicolumn{2}{c}{CN-Celeb2}   \\
    \cmidrule(r){2-3}    \cmidrule(r){4-5}
                              &  \# of Utters    &   Proportion      &  \# of Utters   &   Proportion  \\
    \cmidrule{1-5}
             \textless{2}     & 41,658           &   32.02\%         &  36,505         &   6.89\%    \\
             2-5              & 38,629           &   29.69\%         &  57,215         &   10.81\%   \\
             5-10             & 23,497           &   18.06\%         &  266,799        &   50.39\%   \\
             10-15            & 10,687           &    8.21\%         &  154,120        &   29.11\%   \\
             \textgreater{15} & 15,638           &   12.02\%         &  14,846         &   2.80\%   \\
    \cmidrule{1-5}
  \end{tabular}}
\end{table}


\item CN-Celeb2 includes more data for the genres that were not well covered by CN-Celeb1, such as vlog, live broadcast.
The genre distributions of CN-Celeb1 and CN-Celeb2 are shown in Table~\ref{tab:genrestats}.

\begin{table*}[htb!]
 \begin{center}
  \caption{The genre distribution of \emph{CN-Celeb1} and \emph{CN-Celeb2}.}
   \label{tab:genrestats}
     \begin{tabular}{lllllll}
      \cmidrule{1-7}
   \multirow{2}{*}{Genres} &\multicolumn{3}{c}{CN-Celeb1} & \multicolumn{3}{c}{CN-Celeb2} \\
      \cmidrule(r){2-4}      \cmidrule(r){5-7}
                              &  \# of Spks  &  \# of Utters    &  \# of Hours   &  \# of Spks  &  \# of Utters    &  \# of Hours  \\
      \cmidrule{1-7}
             Advertisement    &  17          &  120             &  0.18          &  66          &   1,542          &   3.86        \\
             Drama            &  160         &  7,247           &  6.43          &  268         &   13,116         &   16.32       \\
             Entertainment    &  483         &  22,064          &  33.67         &  616         &   31,982         &   60.84       \\
             Interview        &  780         &  59,317          &  135.77        &  519         &   34,024         &   81.28       \\
             Live Broadcast   &  129         &  8,747           &  16.35         &  388         &   167,019        &   439.95      \\
             Movie            &  62          &  2,749           &  2.20          &  133         &   4,449          &   5.77        \\
             Play             &  69          &  4,245           &  4.95          &  127         &   14,992         &   22.04       \\
             Recitation       &  41          &  2,747           &  4.98          &  218         &   58,231         &   129.18      \\
             Singing          &  318         &  12,551          &  28.83         &  394         &   42,157         &   75.19       \\
             Speech           &  122         &  8,401           &  36.22         &  394         &   36,680         &   82.58       \\
             Vlog             &  41          &  1,894           &  4.15          &  488         &   125,293        &   177.00      \\
      \cmidrule{1-7}
             Overall          &  1,000       &  130,109         &  273.73        &  2,000       &   529,485        &   1090.01     \\
      \cmidrule{1-7}
     \end{tabular}
 \end{center}
\end{table*}

\item CN-Celeb2 contains less multi-genre speakers than CN-Celeb1.
Fig.~\ref{fig:multi} shows the distribution of multi-genre speakers in the two datasets.
The reason is that the number of celebrities who are active in multiple domains is limited, making it very difficult to collect multi-genre data.
Compared to CN-Celeb1 where 75\% speakers are multi-genre, there are only 33\% multi-genre speakers in CN-Celeb2.
Note that although a large proportion of the speakers are single-genre, the speech utterances were still collected from multiple
sessions in diverse conditions.
Such multi-session data is also very valuable and can be used to develop techniques that can address the
multi-genre challenge with limited multi-genre data.

\begin{figure}[htbp]
\centering
\includegraphics[width=0.9\linewidth]{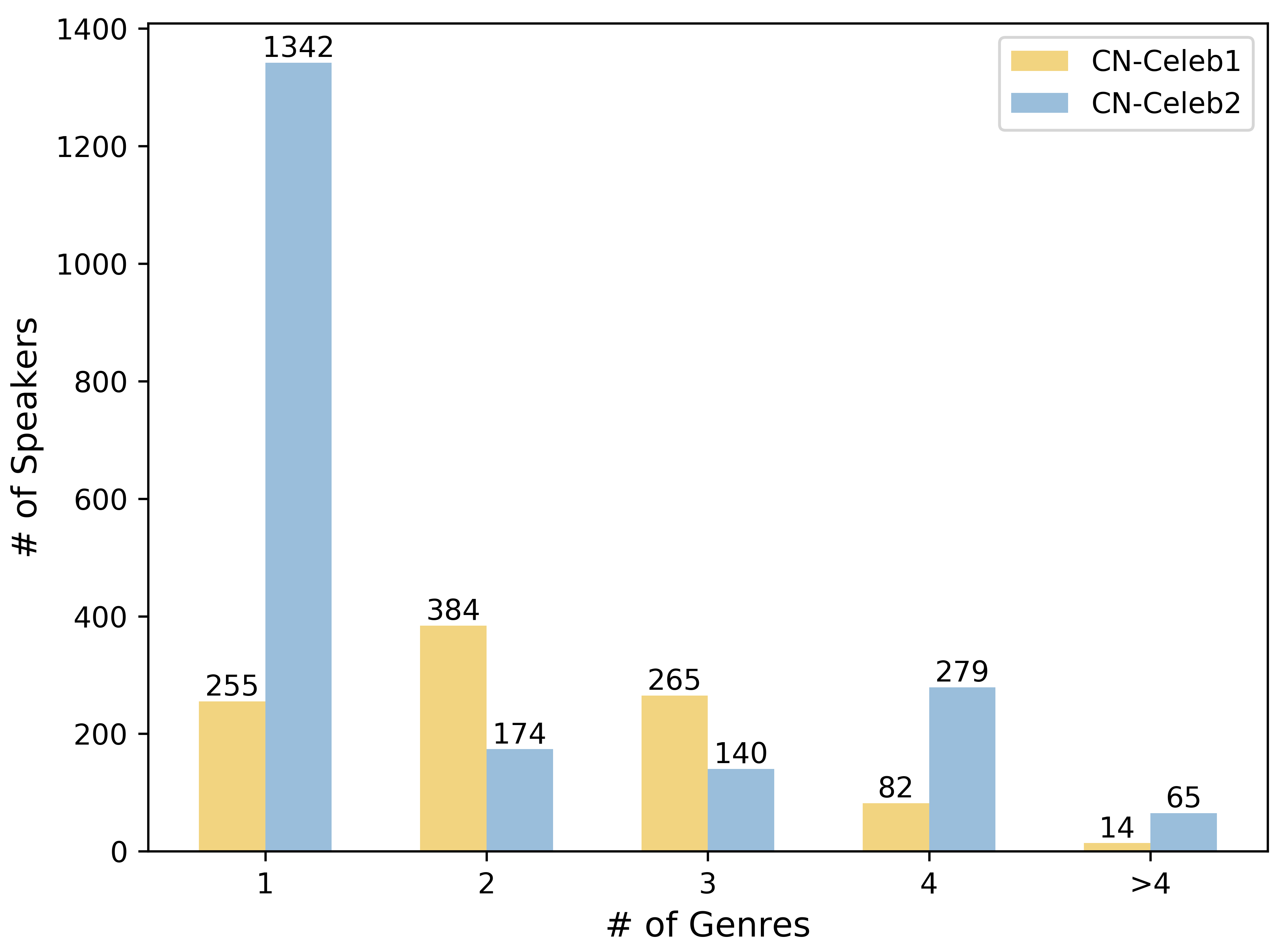}
\caption{The distribution of multi-genre speakers in \emph{CN-Celeb1} and \emph{CN-Celeb2}.}
\label{fig:multi}
\end{figure}

\item The averaged number of utterances per speaker is $265$ in CN-Celeb2 and $130$ in CN-Celeb1.
Fig.~\ref{fig:utter} shows the number of speakers  that have different numbers of utterances in the two datasets, where
the minimum length of the utterances counted in the statistics is set to be $0$s, $2$s, $5$s, $10$s respectively in the four plots.
In other words, utterances shorter than the minimum length are ignored when computing the statistics.
It reflects the proportion of speakers that have different numbers of utterances.

\begin{figure}[htbp]
\centering
\includegraphics[width=1\linewidth]{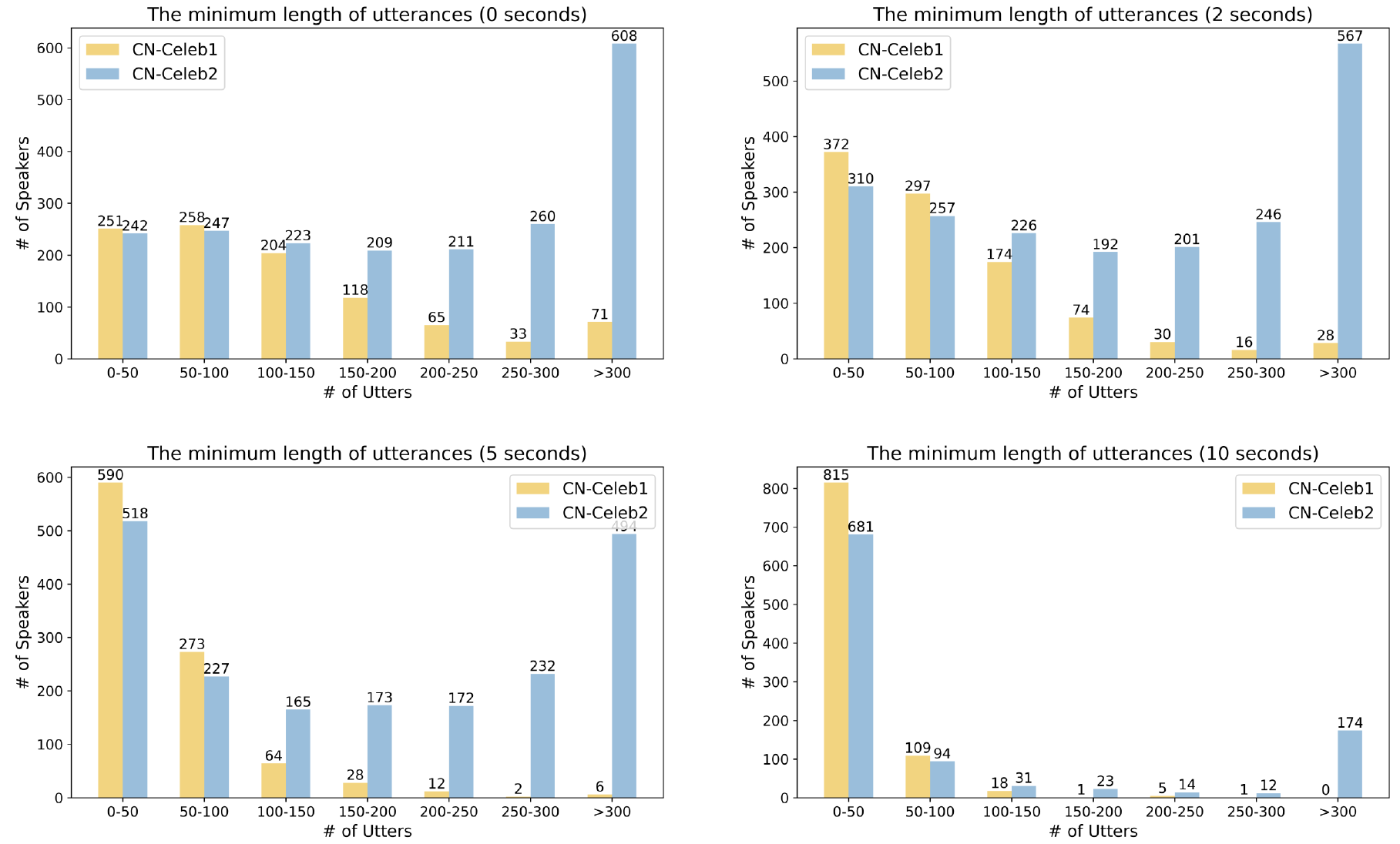}
\caption{The distribution of speakers that have different numbers of utterances in \emph{CN-Celeb1} and \emph{CN-Celeb2}.
For the four plots, the minimum length of the utterances is set to be $0$s, $2$s, $5$s, $10$s respectively, representing that
utterances shorter than the minimum length are ignored when computing the statistics. }
\label{fig:utter}
\end{figure}

\item The averaged number of sessions per speaker is $17$ in CN-Celeb2 and $6$ in CN-Celeb1. Note that we treat each
video as a single session, even though some videos may involve multiple sessions (e.g., in a movie or play).
Fig.~\ref{fig:sess} shows the number of speakers  that have different numbers of sessions in the two datasets, where
the minimum length of the utterances counted in the statistics is set to be $0$s, $2$s, $5$s, $10$s respectively in the four plots.
It reflects the proportion of speakers that have different numbers of sessions.

\begin{figure}[htbp]
\centering
\includegraphics[width=1\linewidth]{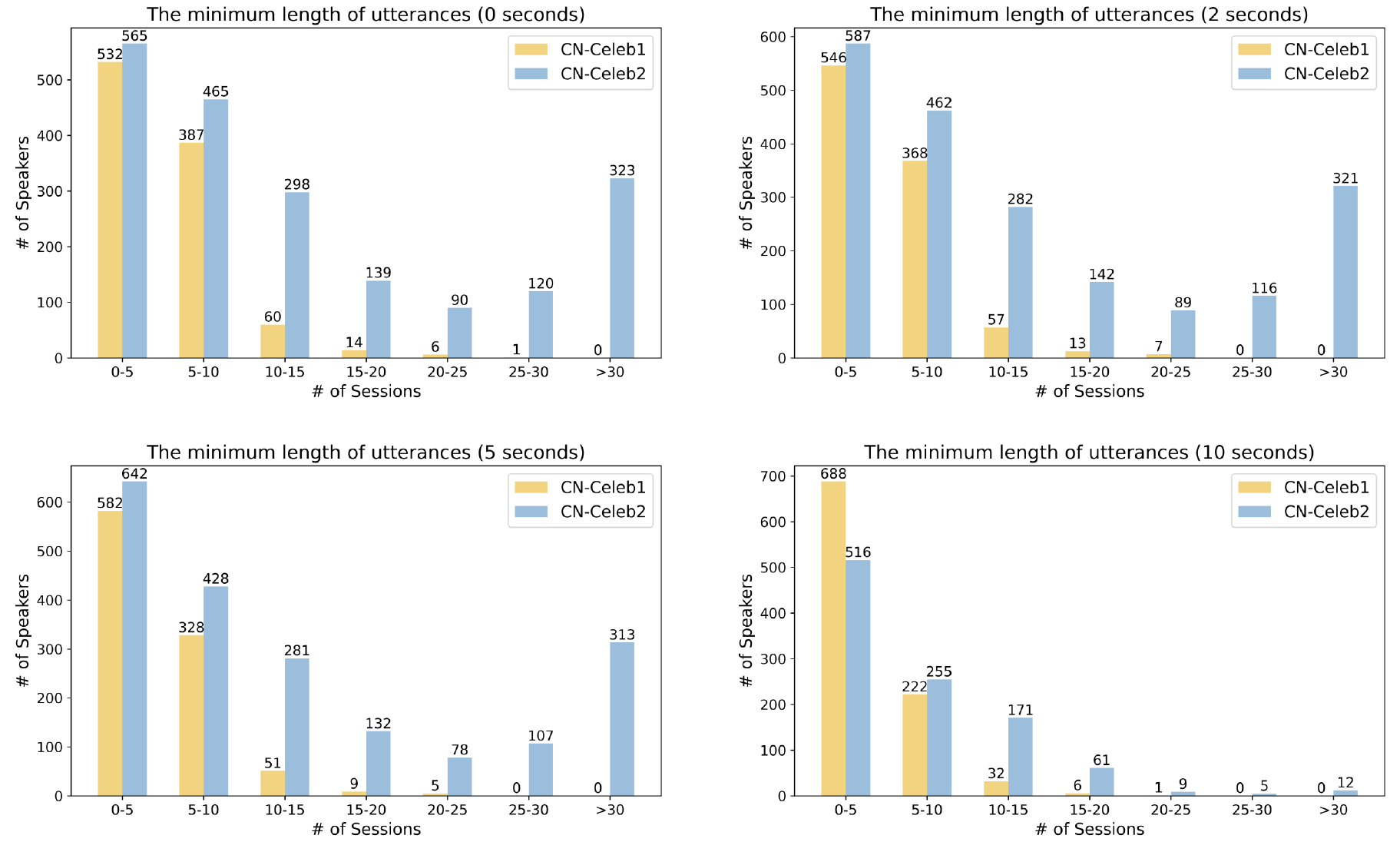}
\caption{The distribution of speakers that have different number of sessions in \emph{CN-Celeb1} and \emph{CN-Celeb2}.
For the four plots, the minimum length of the utterances is set to be $0$s, $2$s, $5$s, $10$s respectively, representing that
utterances shorter than the minimum length are ignored when computing the statistics.}
\label{fig:sess}
\end{figure}

\end{itemize}

We also compare CN-Celeb1 and CN-Celeb2 with some existing datasets in Table~\ref{tab:datasets}.
Note that NIST SRE dataset has not been enumerated as it is not a standalone corpus and changes in composition with each release.

\begin{table*}[htb!]
 \begin{center}
  \caption{Comparison of existing speaker recognition datasets.}
   \label{tab:datasets}
   \centering
    \scalebox{0.92}{
     \begin{tabular}{llllllll}
       \cmidrule{1-7}
         \textbf{Name}   &\textbf{Collection Environment}    &\textbf{Language}   &\textbf{Data Source}   &\textbf{\# of Spks}   &\textbf{\# of Utters} & \textbf{Free}\\
       \cmidrule{1-7}
         Forensic Comparison~\cite{Forensicdatabase}   & clean         & Australian English   & mobile            & $552$      & $1,264$       & Yes   \\
         Free ST Chinese Mandarin\footnotemark[1]      & clean         & Chinese              & mobile            & $855$      & $102,600$     & Yes   \\
         TIMIT~\cite{fisher1986ther,zue1996transcription}                   & clean         & English              & telephone         & $630$      & $6,300$       & No    \\
         SWB~\cite{godf1992swi}                        & clean         & English              & telephone         & $3,114$    & $33,039$      & No    \\
         CSLU\footnotemark[2]     & mostly clean  & English              & telephone         & $500$      & $6,000$      & No    \\
         NIST SRE ~\cite{gonzalez2014evaluating,greenberg2020two}  & clean, noisy  & Multilingual  & telephone, microphone  & $-$     & $-$    & No  \\
         Aishell-1~\cite{bu2017aishell}              & clean         & Chinese              & mobile            & $400$        & $140,000$    & Yes   \\
         Aishell-2~\cite{du2018aishell}              & clean         & Chinese              & mobile            & $1,991$      & $1,000,000$  & Yes   \\
         RSR2015~\cite{lar2014text-dep}                & clean         & English              & mobile, tablet    & $300$        & $190,000$    & No    \\
         RedDots~\cite{lee2015reddots}\footnotemark[3] & clean         & Multilingual         & mobile            & $62$         & $13,500$      & Yes  \\
         HI-MIA~\cite{himia}                           & near/far-field    & Chinese, English & microphone, mobile & $340$     & $3,940,000$   & Yes   \\
         BookTubeSpeech~\cite{Towardbetter}            & multi-media       & English          & BookTube           & $8,450$    & $38,707$       & Yes   \\
         SITW~\cite{mclaren2016speakers}               & interview         & English          & open-source media  & $299$      & $2,800$       & Yes   \\
         VoxCeleb1~\cite{nagrani2017voxceleb}          & mostly interview  & Mostly English  & YouTube                  & $1,251$    & $153,516$     & Yes   \\
         VoxCeleb2~\cite{chung2018voxceleb2}           & mostly interview  & Multilingual    & YouTube                  & $6,112$    & $1,128,246$   & Yes   \\
         CN-Celeb1~\cite{fan2020cn}                    & multi-genre        & Chinese         & Bilibili                 & $1,000$    & $130,109$     & Yes   \\
         \textbf{CN-Celeb2}   & \textbf{multi-genre} & \textbf{Chinese}  & \textbf{multi-media sources}  &  $\textbf{2,000}$  &  $\textbf{529,485}$  & Yes   \\
       \cmidrule{1-7}
     \end{tabular}}
 \end{center}
\end{table*}
\footnotetext[1]{http://www.openslr.org/38/}
\footnotetext[2]{https://catalog.ldc.upenn.edu/LDC2006S26}
\footnotetext[3]{Here presents RedDots\_r2015q4\_v1 released up to August 17th 2015.}

\subsubsection{Collection pipeline}

CN-Celeb2 was collected following a similar pipeline as CN-Celeb1.
The source code has been published online to help readers reproduce our work and collect their own data\footnote{https://github.com/celebrity-audio-collection/videoprocess}.

Broadly, the collection process comprises two stages: in the first stage,
potential segments of the Person of Interest (POI) were extracted from a large amount of raw videos
with an automatic tool, and then in the second stage, human check was employed to remove incorrect segments.
This process is much faster than purely human-based segmentation,
and also avoids potential errors caused by a purely automated process, as VoxCeleb has employed.
We highlight that the human check is important in our case:
because the multi-genre data are very complex in both video and audio, the purely automatic
process cannot deal with such complexity. Although the human check makes the process more costly, it results in a more valuable dataset.

Our automatic pipeline is largely borrowed from the one used for
collecting VoxCeleb1~\cite{nagrani2017voxceleb} and VoxCeleb2~\cite{chung2018voxceleb2},
with some modifications to increase the efficiency and precision.
In particular, we introduced an additional relaxation \& recovery step that employs both image and speech information to validate the extracted segments.
The detailed steps of the collection process are summarized as follows.

\begin{itemize}

\item \textbf{STEP 1. POI list design}.
We manually selected $2,000$ Chinese celebrities as our target speakers. These speakers were mostly from the
entertainment sector, including singers, drama actors/actresses, news reporters and interviewers.
Regional diversity was also taken into account so that variations in accent were covered.

\item \textbf{STEP 2. Pictures and videos download}.
Pictures and videos of the $2,000$ POIs were downloaded from several media sources by searching
for the names of the persons.


For pictures, we developed a crawler to download pictures of POIs using the search engine Baidu\footnote{https://image.baidu.com/}.
For each POI,
120 pictures were downloaded and 10 clear pictures were selected by a human examiner.
Since the POIs are well-known, the selection was very easy and the errors were rare.
We then arranged a double check process, in which the selected 10 pictures were rechecked by another examiner.

For videos, we firstly searched the POI name in the source media.
In order to specify that we were searching for POI names, the word `human' was appended to the search queries.
Secondly, for each POI, at most 10 videos were manually selected and downloaded for each genre, depending on how many
videos can be found in that genre. One examiner was responsible for one POI, and a double check process was
arranged to ensure the quality.

\item \textbf{STEP 3. Face detection and tracking}.
For each POI, we firstly obtained the portrait of the person by detecting and clipping the face images
from all pictures of that person.
The RetinaFace algorithm was used for the detection and clipping~\cite{deng2019retinaface}.
Thereafter, video segments that contained the target person were extracted. This was achieved via the following three-step process: (1) For each frame, detect all the faces appearing in the frame using RetinaFace;
(2) Determine if the target person appears by comparing the POI portrait and the faces detected in the frame using the ArcFace face recognition system~\cite{deng2019arcface}.
(3) Apply the MOSSE face tracking system~\cite{bolme2010visual} supported by OpenCV Tracker\footnote{https://learnopencv.com/object-tracking-using-opencv-cpp-python/}
to produce face streams.


\item \textbf{STEP 4. POI speaking verification by SyncNet}.
As in~\cite{nagrani2017voxceleb}, we employed a mouth-speech synchronization detection system~\cite{chung2016out} to verify
that the target person (POI) is speaking, by testing if the mouth movement of the target person is
synchronized with the speech signal.
This is necessary especially in genres such as movie, drama and entertainment where
the target person appears in the video but the speech is
from other persons. A pre-trained SyncNet model\footnote{\url{https://github.com/joonson/syncnet_python}} was used in our implementation.


\item \textbf{STEP 5. Relaxation \& recovery by speaker diarization}.
Although SyncNet worked well in most cases, it failed for videos of complex genres such as advertisement, movie and vlog.
In these genres, scenes may change abruptly in time, leading to a large number of small POI segments. To solve this problem,
we employed a relaxation \& recheck process: Firstly relaxes the result of SyncNet by merging the adjacent POI segments if their distance is less than $10$
frames, and secondly recovers the true POI part from the merged segments, by using a speaker diarization system.

The details of the recovery step are as follows: we used an off-the-shelf speaker diarization system\footnote{\url{https://github.com/taylorlu/Speaker-Diarization}}
to split the input speech into speaker-homogeneous trunks. The diarization system was based on the UIS-RNN model~\cite{zhang2019fully}, which
firstly split the entire speech into small pieces of $1$s length with $0.6$s overlap,
and then extracted the speaker embeddings of all the pieces using a pre-trained VGG speaker recognition model~\cite{xie2019utterance}.
The UIS-RNN model then clustered the embeddings of all the pieces into several clusters, each corresponding to a single speaker.
According to the clustering result, adjacent pieces were merged together if they were from the same cluster, resulting into
speaker homogeneous trunks.

In order to utilize the diarization result, the \emph{POI cluster} was firstly identified as the one that overlapped with the SyncNet output most.
Then the overlap between the speaker-homogeneous trunks of the POI cluster (output from the diarization) and the merged POI segments from SyncNet (output from the relaxation)
was output as the final POI speech segments.


\item \textbf{STEP 6. Human check}.
The POI segments produced with the above automated pipeline were finally checked by humans.
To ensure the quality, we designed an iterative process: For each POI,
the extracted POI segments were firstly assigned to an examiner for the first-round full check, and then
$20$ segments were sampled and assigned to another examiner for spotting check.
If the accuracy returned by the spotting check was lower than 90\%, the task would be bounced back
to the first examiner for the second-round full check.
This process repeated until the spotting check returned an accuracy higher than 90\%.

According to our experience, this human check is rather efficient: one could check $1$ hour of speech in $1$ hour.
As a comparison, if we do not apply the automated pre-selection, checking $1$ hour of speech requires about $4$ hours.

\end{itemize}

\subsection{Pruning rate}

Human check is the most costly step in the CN-Celeb pipeline. An interesting question is that if this check is necessary, especially for
the genres that are relatively easy to deal with.
For example, for interview, the accuracy of the automatic process might be sufficiently high -- at
least VoxCeleb1 and VoxCeleb2 were collected using a similar pipeline, without any human check.

\begin{figure*}[htbp!]
\centering
\includegraphics[width=0.95\linewidth]{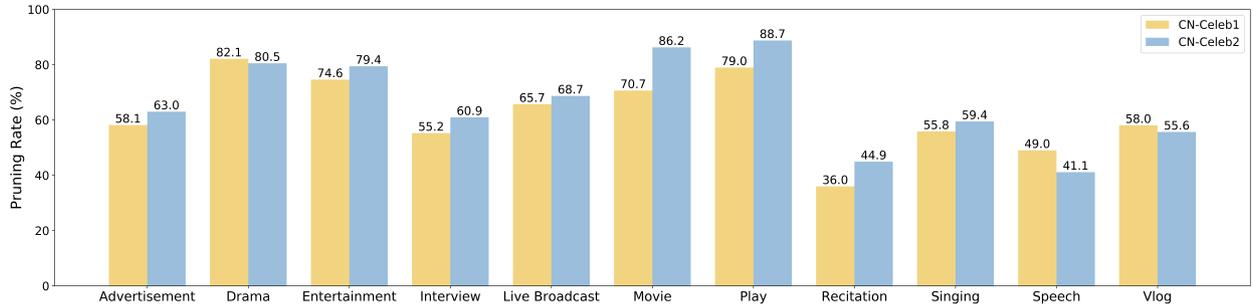}
\caption{Pruning rate with human check when collecting \emph{CN-Celeb1} and \emph{CN-Celeb2}.}
\label{fig:prune}
\end{figure*}

To investigate how necessary the human check is, we compute the \emph{pruning rate} for each genre, i.e., the proportion of frames
that were pruned by human check. The results are shown in Figure~\ref{fig:prune}.
Firstly, it can be seen that for some genres (e.g., speech, recitation, interview), the pruning rate is relatively small, indicating that the automatic process can be regarded as  reliable;
however for other genres (e.g., play, movie), the pruning rate is very high, which means there is a big proportion of frames produced by the automated process that are incorrect.
These results confirm the necessity of the human check in complex genres.
As we will see in the next section (Table~\ref{tab:genre}), speaker recognition performance is often worse on genres with a high pruning rate.
This observation indicates that more complex the genre is, more necessary the human check is.

\section{Experiment I: Multi-genre challenge}

In this section, we will study the basic performance of speaker recognition systems on multi-genre conditions.
Our goal is to investigate the behavior of the state-of-the-art systems when the enrollment/test genre is
different from that of the training, and when the enrollment and test are in different genres.

\subsection{Basic results}

We built two speaker recognition systems, one is based on the i-vector model and one is based on the x-vector model.
The two systems are used as baseline systems to test the single-genre performance and
multi-genre performance on SITW and CN-Celeb.E, respectively.

\subsubsection{Data}

\noindent \textbf{VoxCeleb}\footnote{http://www.robots.ox.ac.uk/$\sim$vgg/data/voxceleb/}: This is used as the training data. It comprises VoxCeleb1 and VoxCeleb2,
amounting to $2,000$+ hours of speech signals from $7,000$+ speakers.
Data augmentation was applied to improve robustness, with the MUSAN corpus~\cite{musan2015} to generate noisy utterances,
and the room impulse responses (RIRS) corpus~\cite{ko2017study} to generate reverberant utterances.

\noindent \textbf{SITW}: This dataset is used for testing single-genre performance.
It comprises $6,445$ utterances from $299$ speakers (precisely, this is the Eval.Core set within SITW).
Note that this dataset is similar to VoxCeleb in terms of data properties, and so there is no
genre mismatch when used as the test set. The test protocol (trials for test) follows the Kaldi SITW recipe\footnote{https://github.com/kaldi-asr/kaldi/egs/sitw}.

\noindent \textbf{CN-Celeb.E}: This dataset is a subset of CN-Celeb1, containing $18,224$ utterances from $200$ speakers.
All the speakers are multi-genre. Note that data of the interview genre is similar to VoxCeleb and SITW, although  they are
from different media sources. Therefore, there is a domain mismatch between training and enrollment/test, while the genres are the same.
During the test, speakers enroll once and are tested against multiple utterances. The enrollment speech (might be split into multiple utterances) for each speaker
is 28s on average.
The average length of the test utterances is 8s, and there are 18,024 test utterances in total, 90 utterances per speaker in average.
The trials are produced by cross pairing the enrollment speech and the test utterances, amounting to 3,604,800 gender-independent test trials.
More details of the test data and trials are shown in Table~\ref{tab:data}.

\begin{table}[htb]
  \caption{Data profile of CN-Celeb.E.}
  \label{tab:data}
   \centering
    \scalebox{1}{
     \begin{tabular}{lll}
       \toprule
        \multirow{2}{*}{Enroll Data}     & Avg. Length per Utt  & 28s \\
                                    & \# of Utters per Spk  & 5 \\
       \midrule
        \multirow{2}{*}{Test Data}       & Avg. Length per Utt  & 8s \\
                                    & \# of Utters per Spk  & 90 \\
       \midrule
        \multirow{2}{*}{Gender Info}     & \# of Female    & 84 \\
                                    & \# of Male      & 116 \\
       \midrule
        \multirow{2}{*}{Trials}     & \# of Target    & 18,024 \\
                                    & \# of Nontarget & 3,586,776 \\
       \bottomrule
     \end{tabular}}
\end{table}

\noindent \textbf{SITW(S)}: This is an auxiliary dataset for performance analysis.
As the average length of SITW is much longer than that of CN-Celeb.E, the results on these two datasets
are not directly comparable. For a more reasonable comparison, we trim the utterances of SITW to match the
average length of the utterances in CN-Celeb.E, which is 28s and 8s for enrollment and test, respectively.
This new dataset is called SITW(S), and the test protocol on this dataset is the same as in SITW.

\subsubsection{Baseline Systems}
In this study, we firstly use the SITW recipe of the Kaldi toolkit~\cite{povey2011kaldi} to build our i-vector and x-vector baselines.
This basic recipe may not achieve the best performance on a particular dataset, but has been demonstrated to be highly competitive and generalizable by many researchers
with their own data and model settings. We therefore consider that this recipe can represent a stable state-of-the-art technique.
Moreover, using this recipe allows others to reproduce our results easily.

\noindent \textbf{i-vector system}:
The i-vector model was built following the Kaldi SITW/v1 recipe.
The acoustic features comprise 24-dimensional MFCCs plus the log energy, augmented by first- and second-order derivatives, resulting in a 75-dimensional feature vector.
Moreover, cepstral mean normalization (CMN) is employed to normalize the channel effect,
and an energy-based voice active detection (VAD) is used to remove silence segments.
The universal background model (UBM) consists of 2,048 Gaussian components, and the dimensionality of the i-vector is set to be 400.
For the back-end model, LDA is firstly employed to reduce the dimensionality to 150, and then PLDA is used to score the trials.

\noindent \textbf{x-vector system}:
The x-vector model was created following the Kaldi SITW/v2 recipe. The acoustic features are 30-dimensional MFCCs.
The DNN architecture involves 5 time-delay (TD) layers to learn frame-level deep speaker features,
and a temporal statistic pooling (TSP) layer is used to accumulate the frame-level features to utterance-level statistics, including the mean and standard deviation.
After the pooling layer, 2 fully-connection (FC) layers are used as the classifier, for which the outputs correspond to
the number of speakers in the training set.
Once trained, the 512-dimensional activations of the penultimate layer are read out as an x-vector.
The back-end model is the same as in the i-vector system, which includes LDA for dimensional reduction, and PLDA to score the trials.

\subsubsection{Baseline results}

The overall results in terms of equal error rate (EER) are shown in Table~\ref{tab:baseline}.
It can be seen that both the i-vector and x-vector systems obtain reasonable performance on SITW, and the results
are similar to the official results released with Kaldi recipes.
The results on SITW(S) are slightly worse than those on SITW, which is expected as the enrollment and test utterances are shorter in this dataset.
The performance on CN-Celeb.E is much worse.
For example, compared to the x-vector results on CN-Celeb.E and SITW(S), the EER on CN-Celeb.E increases by more than 300\%.
These results clearly indicate that the state-of-the-art speaker recognition systems cannot inherently deal with the complexity introduced by multiple genres.

\begin{table}[htb]
  \caption{EER(\%) results with the i-vector and x-vector baseline systems.}
  \label{tab:baseline}
   \centering
    \scalebox{0.88}{
     \begin{tabular}{cccccc}
       \cmidrule{1-6}
       \multirow{3}{*}{System}  &  \multicolumn{2}{c}{Training Set}  &  \multicolumn{3}{c}{Test Set}  \\
                                   \cmidrule(r){2-3}                 \cmidrule(r){4-6}
                                &  Front-end  &  Back-end  &  SITW  &  SITW(S)  &  CN-Celeb.E        \\
       \cmidrule(r){1-1} \cmidrule(r){2-3} \cmidrule(r){4-6}
            i-vector            &  VoxCeleb   &  VoxCeleb  &  5.66  &  7.41     &  18.37     \\
            x-vector            &  VoxCeleb   &  VoxCeleb  &  3.48  &  4.62     &  16.59     \\
       \cmidrule{1-6}
     \end{tabular}}
\end{table}

\subsubsection{More powerful x-vector systems}

We have also implemented more powerful x-vector systems by using arguably more advanced techniques.
In this work, we used an open-source code\footnote{https://github.com/kjw11/tf-kaldi-speaker} and
tested a bunch of state-of-the-art architectures/techniques,
such as ResNet~\cite{chung2018voxceleb2,zeinali2019but},
self-attentive pooling~\cite{zhu2018self} and additive angular margin loss~\cite{deng2019arcface}, on SITW and CN-Celeb.E.
The results are shown in Table~\ref{tab:sota}.
It can be found that although these more advanced systems obtain obvious performance improvements over the basic x-vector system on SITW (1.96\% vs. 3.48\%),
they did not show clear superiority on CN-Celeb.E (16.51\% vs. 16.59\%).
It indicates that these advanced techniques may simply overfit to the training condition and so are of little help in solving the multi-genre challenge.
This is another reason why we use the basic x-vector architecture as our baseline.

\begin{table}[htb]
  \caption{EER (\%) results of more powerful x-vector systems. `TSP' represents temporal statistic pooling.
           `SAP' represents self-attentive pooling. `AAM' represents additive angular margin.}
   \label{tab:sota}
   \centering
    \scalebox{0.96}{
     \begin{tabular}{lllll}
       \cmidrule{1-5}
        Topology       &  Pooling   & Loss           & SITW      & CN-Celeb.E    \\
       \cmidrule{1-5}
        TDNN           &  TSP       & Softmax        & 2.43      & 16.87          \\
        TDNN           &  TSP       & AAM-Softmax    & 2.49      & 16.65          \\
        TDNN           &  SAP       & Softmax        & 2.41      & 17.11          \\
        TDNN           &  SAP       & AAM-Softmax    & 2.57      & 16.96          \\
        ResNet-34      &  TSP       & Softmax        & 2.41      & 16.74          \\
        ResNet-34      &  TSP       & AAM-Softmax    & 1.96      & 16.51          \\
        ResNet-34      &  SAP       & Softmax        & 2.16      & 17.33          \\
        ResNet-34      &  SAP       & AAM-Softmax    & 2.30      & 16.52          \\
       \cmidrule{1-5}
     \end{tabular}}
\end{table}

\subsection{Within-genre results}

\begin{figure*}[b]
  \centering
  \subfigure[i-vector]{\includegraphics[width=0.453\linewidth]{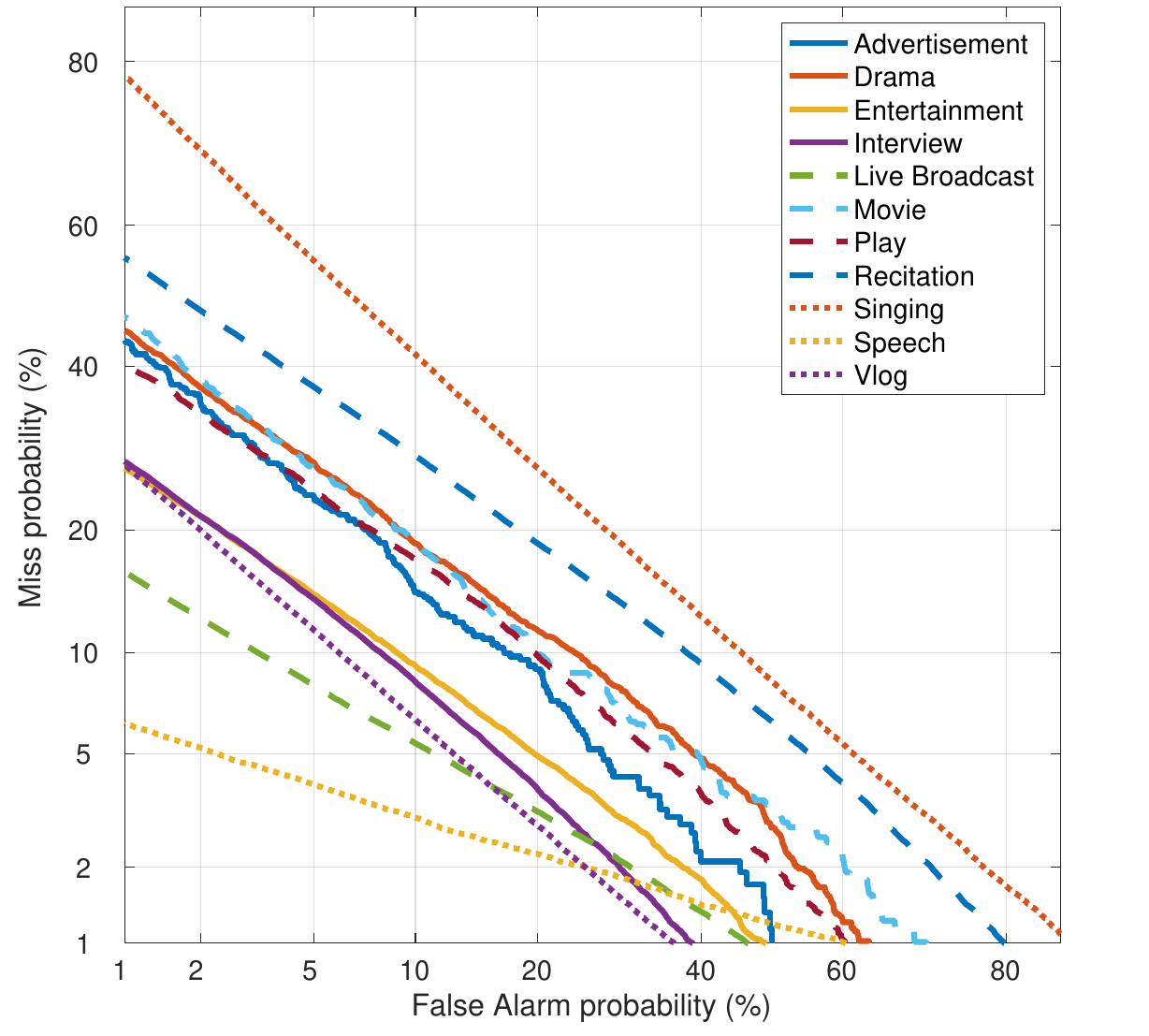}}
  \subfigure[x-vector]{\includegraphics[width=0.438\linewidth]{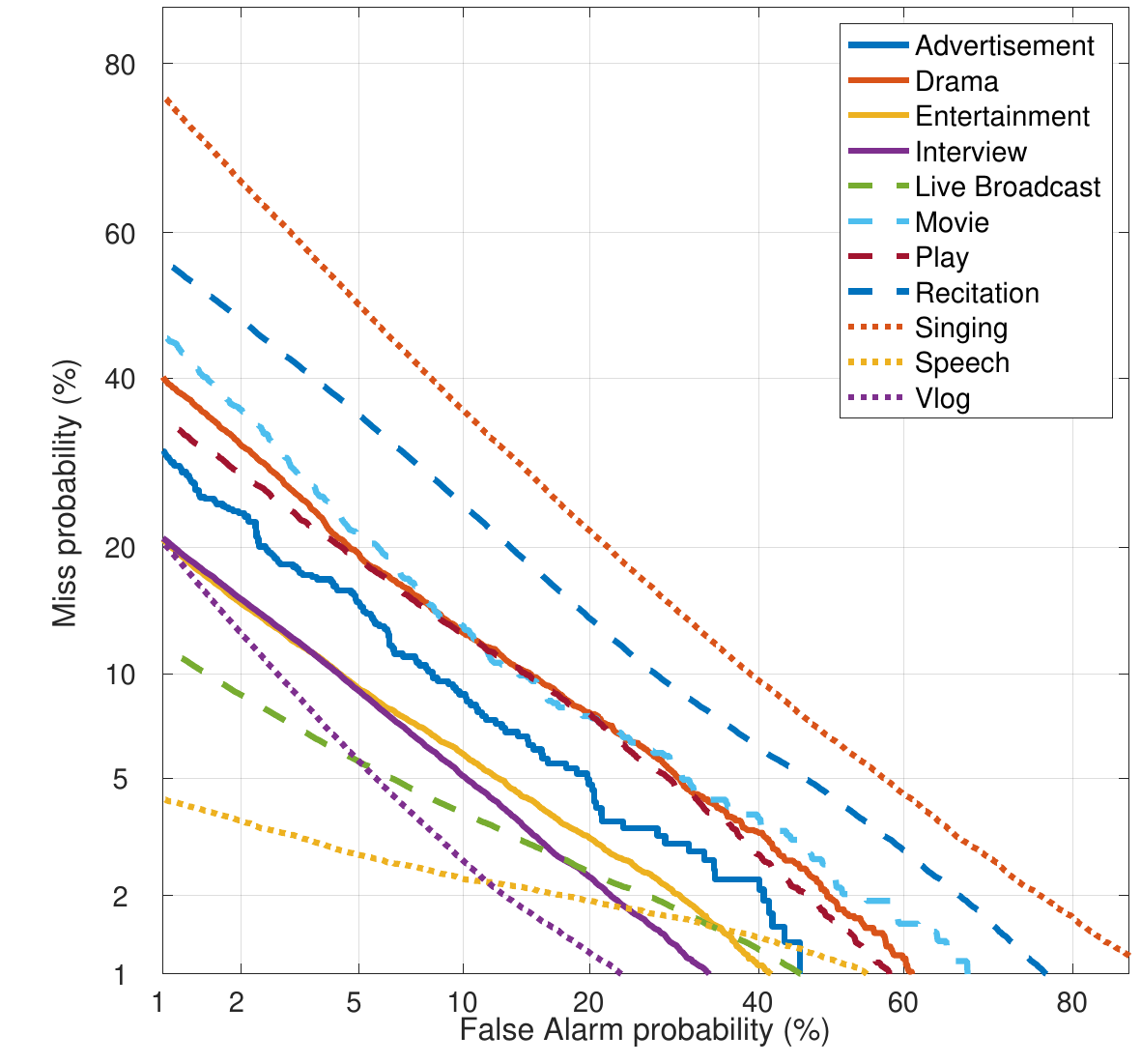}}
  \caption{DET curves of different genres with the i-vector and x-vector systems.}
  \label{fig:det}
\end{figure*}

In this section, we break down the multi-genre tests and compare the performance in different genres. We focus on the case
where the enrollment and test utterances are from the same genre.
To ensure the confidence of experimental results, we use the overall CN-Celeb dataset (3,000 speakers in total) for test
and also filter away short utterances with less than 5s duration.
Table~\ref{tab:genre} presents the EER results, and Figure~\ref{fig:det} shows the DET curves.
In the test for each genre, 5 utterances of each speaker are randomly selected for enrollment, and the remaining utterances are used for test.

\begin{table}[htb!]
  \caption{EER(\%) results of the baseline systems in different genres on CN-Celeb.}
   \label{tab:genre}
   \centering
    \scalebox{0.95}{
     \begin{tabular}{lllll}
      \cmidrule{1-5}
        Genres           &  \# of Spks  &  \# of Utters  &  i-vector  &  x-vector   \\
      \cmidrule{1-5}
        Advertisement    &  75          &  781           &  12.43     &  9.37      \\
        Drama            &  377         &  4,521         &  14.66     &  11.70     \\
        Entertainment    &  1,020       &  18,931        &  9.48      &  7.31      \\
        Interview        &  1,253       &  41,586        &  9.06      &  6.98      \\
        Live Broadcast   &  496         &  154,249       &  6.79      &  5.42      \\
        Movie            &  165         &  1,495         &  14.17     &  11.47     \\
        Play             &  170         &  5,476         &  13.87     &  11.56     \\
        Recitation       &  259         &  58,839        &  19.21     &  16.55     \\
        Singing          &  683         &  38,879        &  23.37     &  20.86     \\
        Speech           &  331         &  39,792        &  4.19      &  3.21      \\
        Vlog             &  524         &  120,812       &  7.92      &  5.31      \\
      \cmidrule{1-5}
        Overall          &  3,000       &  485,361       &  8.75      &  7.43     \\
      \cmidrule{1-5}
     \end{tabular}}
\end{table}

It can be observed that with both the i-vector and the x-vector systems, performance on different genres is substantially different.
For genres such as speech, live broadcast, vlog and interview, the EER results are less than 8\%, and the performance is relatively acceptable.
While for genres such as singing, recitation, drama and movie, the EER results are more than 12\%, and the performance is quite unacceptable.
The performance discrepancy on different genres could be attributed to two reasons:
Firstly, the speaker-independent variation is naturally much more significant for some genres compared to others.
For example, the channel, background and speaking style in speech and interview tend to be more controlled than those in singing and drama.
The more complex the variation, the more difficult it is for the speaker traits to be identified.
Secondly, the i-vector and x-vector models are trained with VoxCeleb that mainly consists of interview speech.
This perhaps makes the model biased to interview and similar genres such as speech and live broadcast.

Another observation is that even for the interview genre, the performance is much worse than that obtained on SITW
(6.98\% vs. 3.48\% on the x-vector system).
This is clearly caused by the discrepancy between channels and languages of the two sources of CN-Celeb and SITW
(Bilibili, etc., for CN-Celeb while YouTube for SITW).
It indicates that the true performance of the present state-of-the-art speaker recognition system is not
as good as one though from the results reported on SITW, even without genre mismatch.

Taking the x-vector system as an example, if we set the overall EER (7.43\%) as the threshold for an \emph{acceptable} system,
the present speaker recognition system can only obtain reasonable performance in a few genres.
These genres are speech, vlog, live broadcast, interview and entertainment.
This clearly demonstrated how challenging the multi-genre problem is.


\subsection{Cross-genre results}

\begin{figure*}[htbp]
  \centering
  \includegraphics[width=1.0\linewidth]{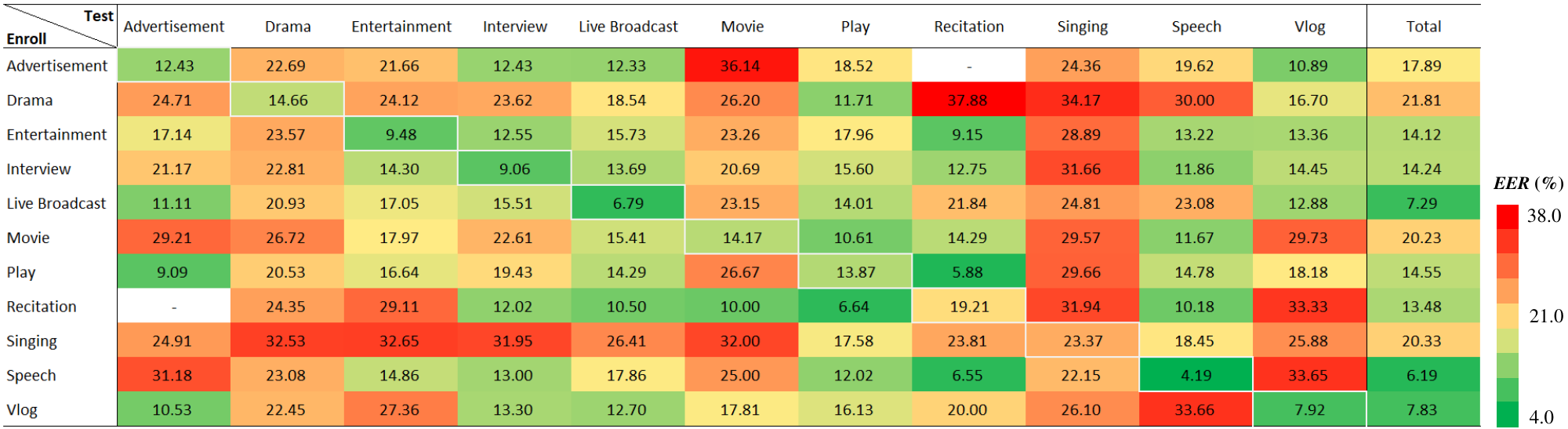}
   \caption{Cross-genre tests with the i-vector system. The lightness of the color corresponds to the numerical value of the EER(\%).}
  \label{fig:matrix-iv}
\end{figure*}

\begin{figure*}[htbp]
  \centering
  \includegraphics[width=1.0\linewidth]{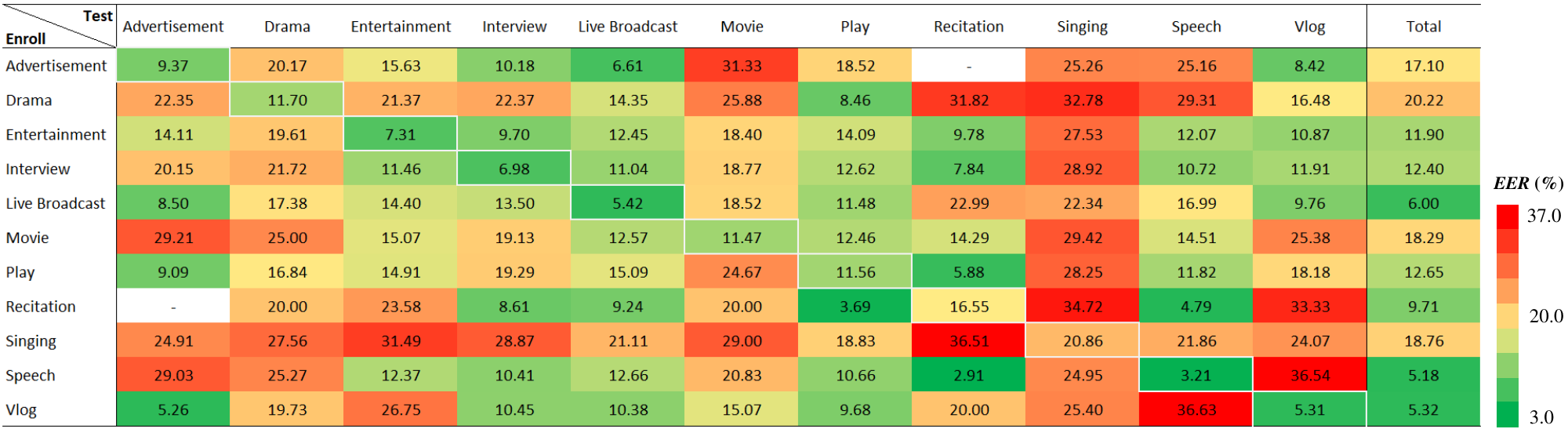}
  \caption{Cross-genre tests with the x-vector system. The lightness of the color corresponds to the numerical value of the EER(\%).}
  \label{fig:matrix-xv}
\end{figure*}

In this section, we focus on cross-genre test and compute a genre-to-genre performance matrix.
Figure~\ref{fig:matrix-iv} and Figure~\ref{fig:matrix-xv} show the EER results with the i-vector system and the x-vector system, respectively.
The numerical values shown in the blocks are the EER results under the enrollment genre corresponding to its row and the test genre corresponding to its column.
Note that the diagonal results show the in-genre results in Table~\ref{tab:genre}.
The last column shows the overall results that the enrollment is based on one genre and test is on all the genres.
Note that there are two blank cells (recitation-advertisement and recitation-advertisement).
This is because there is only 1 recitation-advertisement speaker, which makes the EER result unreliable.

Firstly, paying attention to the overall results (the last column) enrolled with each genre, it can be found that the best performance is obtained when
the enrollment is with the speech genre, and the worst performance is obtained when the enrollment is with the singing genre.
Roughly stating, the simpler the enrollment condition (e.g., speech, interview, etc.), the better is the average performance obtained.
This is good news as one tends to enroll in the silent environments with careful pronunciation.

Secondly, the results with the same enroll-test pair (e.g., singing-speech and speech-signing) are roughly the same.
This phenomenon was also observed in some previous studies on multiple speaking styles, e.g., ~\cite{park2017using,shriberg2009does,park2016speaker,shriberg2008effects}.
This indicates that the variation in the enrollment genre is similar to the variation in the test genre.

Thirdly, the cross-genre performance is determined by two factors:
(1) the complexity of the enrollment/test genre,
(2) the degree of match between the two genres.
For example, when the enrollment is movie, the EER is 14.17\% when the test genre is movie, which is not very bad due to the matched genre; however the
EER is 11.67\% when the test genre is speech, which is even better, due to the simpler condition of the speech genre.
In general, worse performance is obtained when the enrollment and test genres are less matched.

\begin{figure*}[htbp]
  \centering\includegraphics[width=0.9\linewidth]{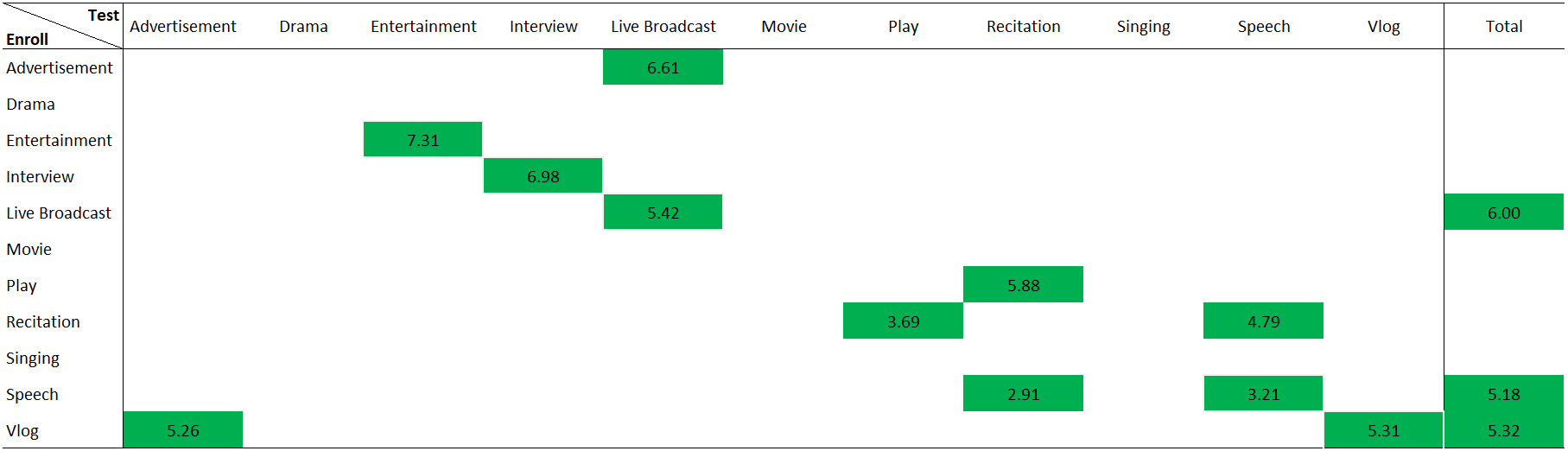}
  \caption{Cross-genre results (EER \%) for acceptable conditions (EER threshold = 7.43\%) with the x-vector baseline system.}
  \label{fig:matrix-xv-thres}
\end{figure*}

In summary, the cross-genre phenomenon is highly complex, and in most of the conditions, the performance is not acceptable.
Taking the x-vector system as an example, if we set the overall EER (7.43\%) as the threshold for acceptance, there are only
several conditions can be deemed acceptable, as shown in Figure~\ref{fig:matrix-xv-thres}.
These observations indicate once again that cross-genre is a very challenging problem.

\begin{figure*}[htbp]
  \centering
  \subfigure[$C_{llr}$ results with the i-vector system under cross-genre tests]{\includegraphics[width=0.95\linewidth]{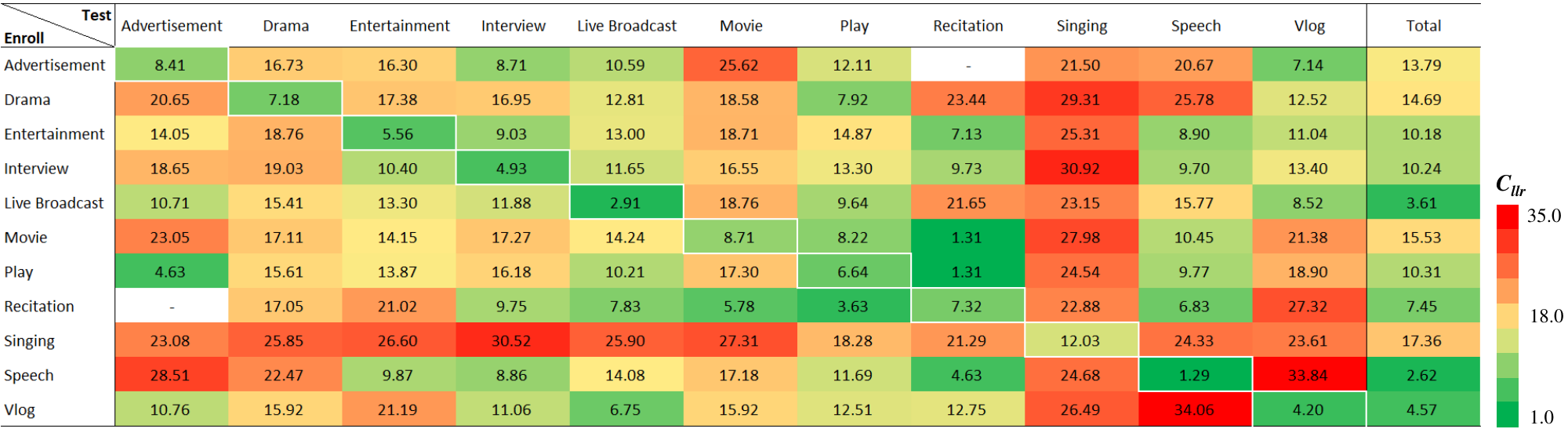}}
  \subfigure[$C_{llr}^{min}$ results with the i-vector system under cross-genre tests]{\includegraphics[width=0.95\linewidth]{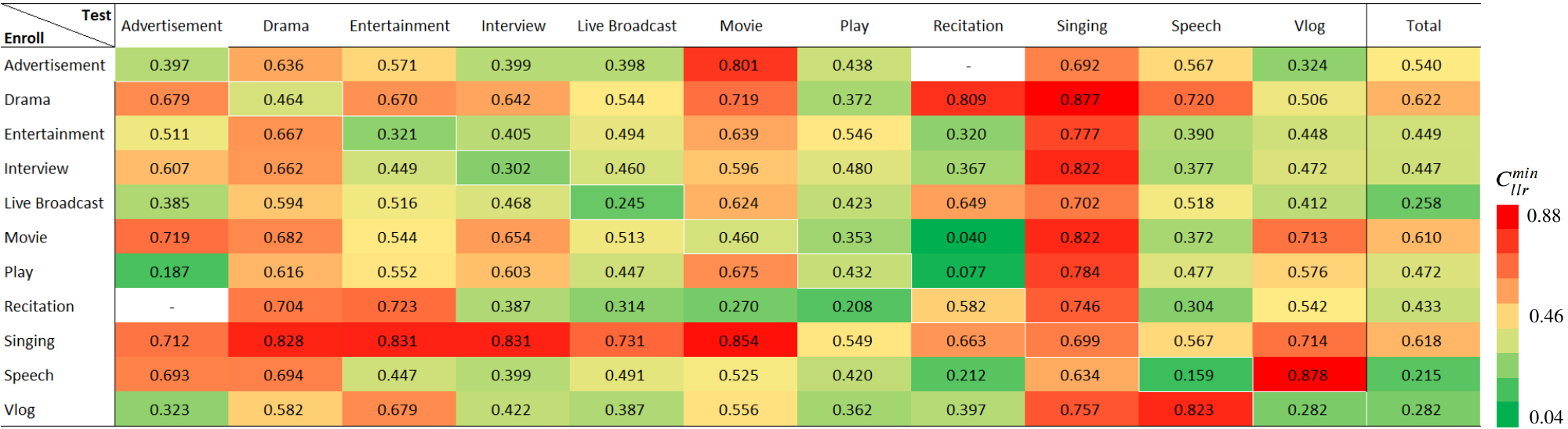}}
   \caption{Cross-genre tests with the i-vector system. The lightness of the color corresponds to the numerical value of the $C_{llr}$/$C_{llr}^{min}$.}
  \label{fig:matrix-iv-cllr}
\end{figure*}

\begin{figure*}[htbp]
  \centering
  \subfigure[$C_{llr}$ results with the x-vector system under cross-genre tests]{\includegraphics[width=0.95\linewidth]{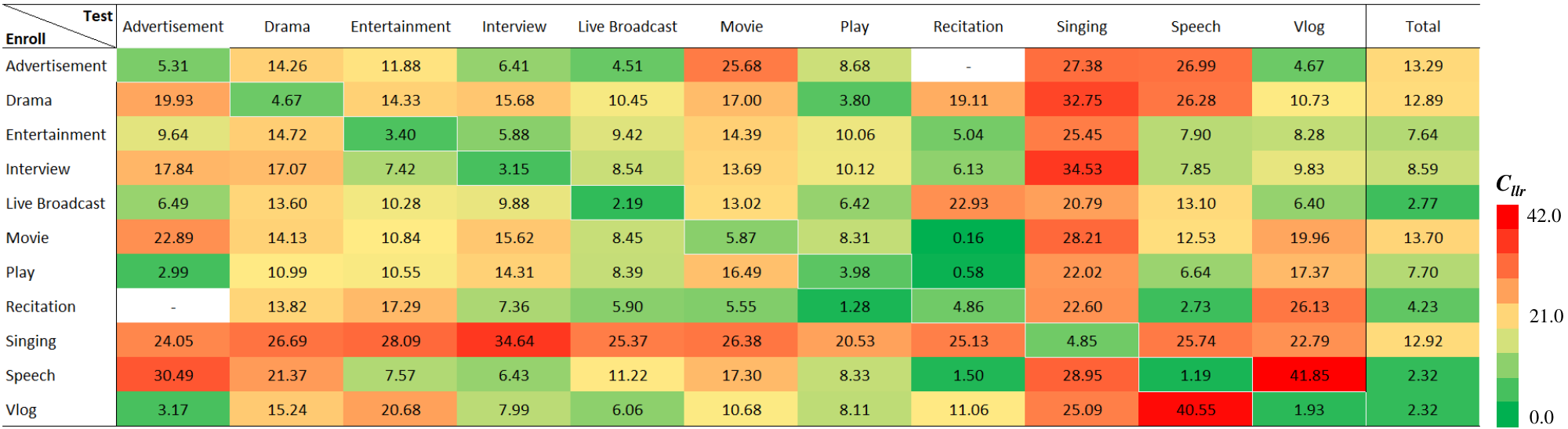}}
  \subfigure[$C_{llr}^{min}$ results with the x-vector system under cross-genre tests]{\includegraphics[width=0.95\linewidth]{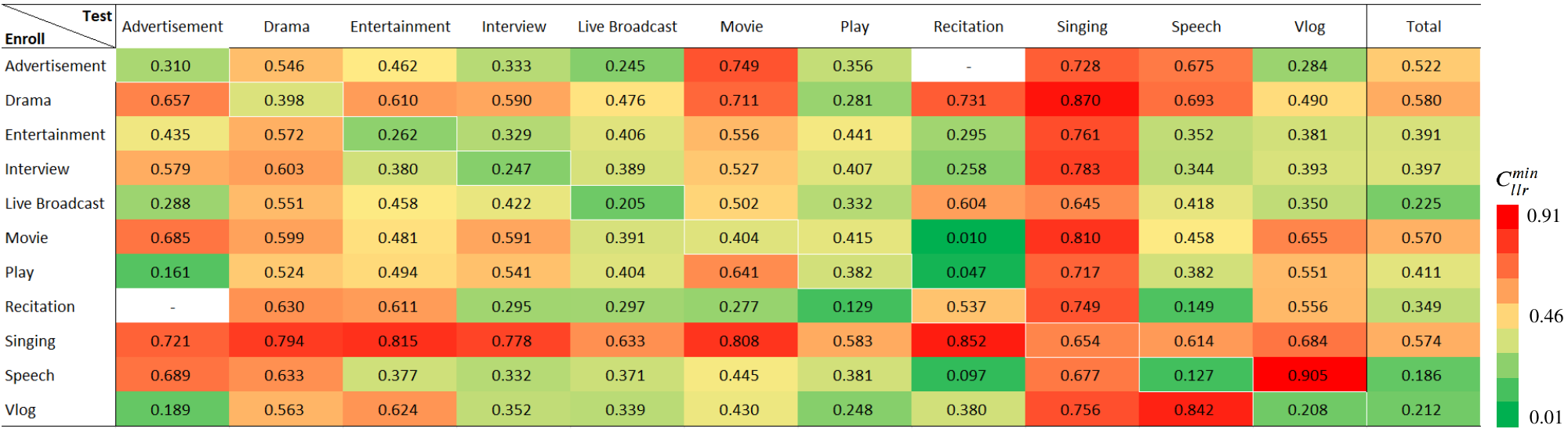}}
   \caption{Cross-genre tests with the x-vector system. The lightness of the color corresponds to the numerical value of the $C_{llr}$/$C_{llr}^{min}$.}
  \label{fig:matrix-xv-cllr}
\end{figure*}

For a deep analysis, we also report results in two metrics related to score calibration: the cost of log likelihood ratio (LLR) denoted by $C_{llr}$ and the minimum cost
of LLR denoted by $C_{llr}^{min}$~\cite{ramos2008cross,brummer2006application}. Compared to EER, $C_{llr}$ evaluates the averaged performance
over all the possible settings on the priors and costs of the target and non-target trials, by assuming that the PLDA scores are LLRs.
In the cross-genre situation, the enrollment and test conditions are drastically different, and so the PLDA scores may significant deviated from
the true LLRs. In this case, a large $C_{llr}$ could be attributed to the lack of both discrimination and regularization of the PLDA scores, where discrimination
refers to how well the scores can distinguish target and non-target trails, while regularization refers to how well the scores represent LLRs .
A score calibration can be designed to map the scores to LLRs. This map is monotonic and so does not change the discrimination capacity of the scores but
improves regularization. When the calibration is perfect (which can be obtained with a finite evaluation set), the resultant $C_{llr}$ is $C_{llr}^{min}$.
Therefore the difference between $C_{llr}$ and $C_{llr}^{min}$ (sometimes called $C_{loss}$ )
reflects how the PLDA scores biased from LLRs and how much the score calibration may contribute.

It should be noted that score calibration does not change EERs with each cross-genre test as the calibration is simply a monotonic score mapping.
However, it may improve system performance on the overall test, by applying different calibration models for different cross-genre tests.
This is because after the test-dependent calibration, the scores of different tests become comparable and a cross-test threshold is applicable.

The $C_{llr}$/$C_{llr}^{min}$ results are shown in Table~\ref{fig:matrix-iv-cllr} and Table~\ref{fig:matrix-xv-cllr} for the i-vector and x-vector systems, respectively.
Note that the last column `Total' reports the results with all the test trials pooled of each row.

We firstly observe that the performance tendency of the $C_{llr}^{min}$ results are similar to that of the EER results.
This is not very surprising as both evaluate the discrimination power of the scores, though $C_{llr}^{min}$ reflects the expected
error rate while EER is the error rate at the equilibrium point of false acceptance and false rejection.
Due to the similar trend, we will keep use EER as the main metric and report the EER results only when discussing the relative performance.

Moreover, we found that there is a large gap between $C_{llr}$ and $C_{llr}^{min}$, and this gap is more significant for the test scenarios where the enroll-test
mismatch is more obvious (specified by results in EER and $C_{llr}^{min}$). This indicates that the PLDA scores are far from LLRs, and score calibration is important
in real applications where a threshold is required for decision making.

\subsection{Statistical analysis}

In this section, we analyze the performance degradation under within-genre and cross-genre conditions.
A key insight is that if the distribution of the speaker vectors remains the same as in the training condition,
then the performance with the PLDA scoring will be optimal assuming the PLDA model is well trained with the training data~\cite{wang2020remarks}.
Therefore, the performance degradation we observe in the multi-genre test (either the within-genre test or the cross-genre test) can be
understood by the change in the statistics of the distribution.
Since PLDA is the back-end scoring model, we compute the statistics related to PLDA, including
the inter-speaker variance, intra-speaker variance and the global mean shift.
We will compute these statistics of each genre and observe the statistics change amongst different genres.

We firstly compute the inter-speaker variance and intra-speaker variance of each genre in Table~\ref{tab:genre-var}.
Besides, we also compute the mean shift between VoxCeleb and different genres of CN-Celeb.
The mean vector of VoxCeleb is regarded as a reference vector,
and the mean vectors of different genres in CN-Celeb are regarded as genre vectors.
The mean shifts can be computed between the reference vector and different genre vectors based on the Euclidean distance and Cosine similarity.
Results are shown in Table~\ref{tab:mean-dist}.
Note that all these results are computed in the PLDA transformed space.


\begin{table}[htp!]
  \caption{Inter-speaker variances ($S_b$) and intra-speaker variances ($S_w$) of i-vectors and x-vectors derived from VoxCeleb and different genres of CN-Celeb. }
  \label{tab:genre-var}
   \centering
    \scalebox{1}{
    \begin{tabular}{lllll}
      \cmidrule{1-5}
       \multirow{2}{*}{Genres}   &  \multicolumn{2}{c}{i-vector}  & \multicolumn{2}{c}{x-vector} \\
      \cmidrule(r){2-3}      \cmidrule(r){4-5}
                      &  $S_b$  & $S_w$      & $S_b$   & $S_w$        \\
      \cmidrule{1-5}
       VoxCeleb       & 0.920 & 1.042    & 1.053 & 0.894       \\
      \cmidrule{1-5}
       Advertisement  & 0.443 & 1.020    & 1.162 & 3.167       \\
       Drama          & 0.297 & 1.227    & 0.982 & 4.835       \\
       Entertainment  & 0.300 & 1.176    & 0.889 & 4.179       \\
       Interview      & 0.297 & 1.114    & 0.755 & 3.967       \\
       Live Broadcast & 0.357 & 1.052    & 0.822 & 2.023       \\
       Movie          & 0.404 & 1.210    & 1.201 & 5.289       \\
       Play           & 0.250 & 1.232    & 0.868 & 4.740       \\
       Recitation     & 0.360 & 1.481    & 0.816 & 2.379       \\
       Singing        & 0.204 & 1.226    & 0.618 & 2.884       \\
       Speech         & 0.542 & 1.096    & 1.038 & 2.225       \\
       Vlog           & 0.349 & 1.307    & 0.830 & 2.726       \\
      \cmidrule{1-5}
    \end{tabular}}
\end{table}


\begin{table}[htb!]
  \caption{Mean shifts of i-vectors and x-vectors on different genres of CN-Celeb. \emph{Euc.} represents Euclidean distance and \emph{1-Cos.} represents Cosine similarity.}
   \label{tab:mean-dist}
   \centering
    \scalebox{1}{
     \begin{tabular}{lllll}
      \cmidrule{1-5}
        \multirow{2}{*}{Genres}   & \multicolumn{2}{c}{i-vector}  & \multicolumn{2}{c}{x-vector} \\
      \cmidrule(r){2-3}      \cmidrule(r){4-5}
                         & \emph{Euc.}  & \emph{1-Cos.}  & \emph{Euc.}   & \emph{1-Cos.}    \\
      \cmidrule{1-5}
        Advertisement    & 1.616 & 1.305 & 1.554 & 1.208   \\
        Drama            & 1.575 & 1.240 & 1.515 & 1.148   \\
        Entertainment    & 1.590 & 1.263 & 1.551 & 1.203   \\
        Interview        & 1.562 & 1.220 & 1.532 & 1.174   \\
        Live Broadcast   & 1.498 & 1.122 & 1.533 & 1.174   \\
        Movie            & 1.587 & 1.259 & 1.519 & 1.153   \\
        Play             & 1.482 & 1.097 & 1.449 & 1.049   \\
        Recitation       & 1.491 & 1.111 & 1.498 & 1.123   \\
        Singing          & 1.653 & 1.366 & 1.526 & 1.164   \\
        Speech           & 1.410 & 0.994 & 1.426 & 1.017   \\
        Vlog             & 1.516 & 1.150 & 1.543 & 1.190   \\
      \cmidrule{1-5}
   \end{tabular}}
\end{table}

Firstly, it can be found that the statistics of VoxCeleb are more similar to matched genres (e.g., speech and interview) compared to unmatched genres (e.g., singing and recitation),
and the mean shift is less significant in the case of matched genres. As the PLDA is trained on VoxCeleb, if the statistics change and the mean shift are
significant, the performance will be impacted. Referring to the results in Table~\ref{tab:genre}, it can be observed that the genre incurring the most significant
statistics change and mean shift suffers from the most performance reduction.

The statistics change and the mean shift cause more severe problems in the cross-genre scenario, as the enroll data and test data in this scenario
possess different statistical properties but they have to be represented in a single PLDA model.
We presented a deep analysis on this enroll-test mismatch problem in our recent study~\cite{li2020aprinciple},
but mismatch caused by the cross-genres challenge is yet to be thoroughly studied.

\subsection{Qualitative analysis}

\begin{figure*}[htbp]
\centering
\includegraphics[width=0.95\linewidth]{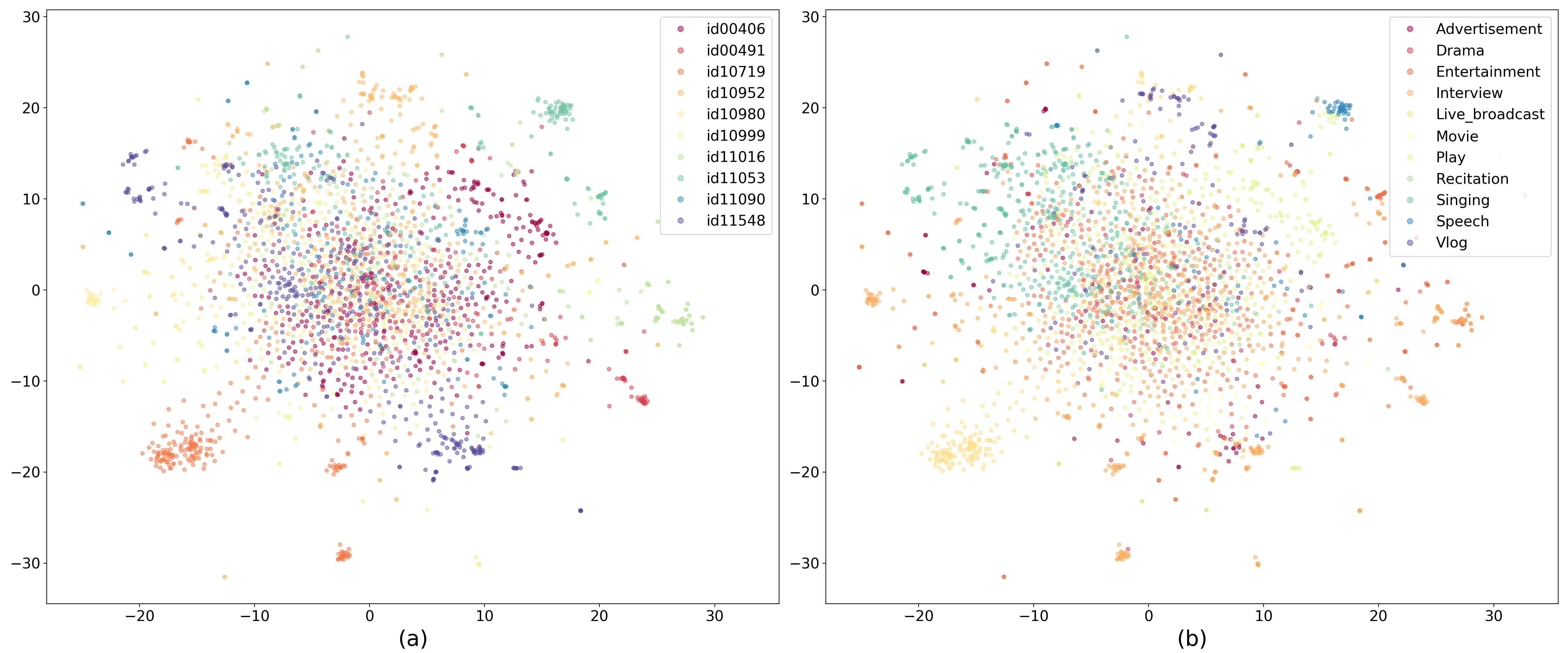}
\caption{The i-vector distribution plotted by t-SNE, where (a) each color represents a speaker, (b) each color represents a genre.}
\label{fig:ti}
\end{figure*}

\begin{figure*}[htbp]
\centering
\includegraphics[width=0.95\linewidth]{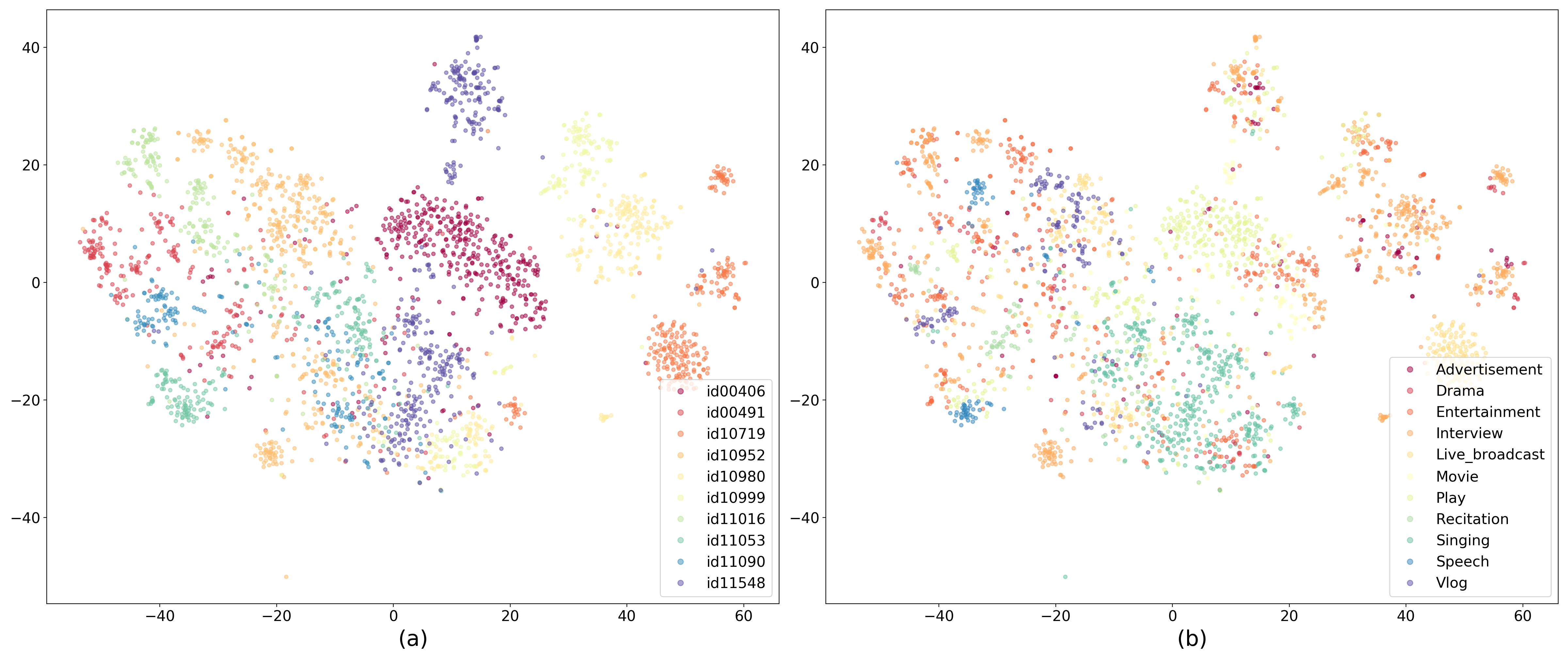}
\caption{The x-vector distribution plotted by t-SNE, where (a) each color represents a speaker, (b) each color represents a genre.}
\label{fig:tx}
\end{figure*}

In this section, we analyze the distribution of the speaker vectors by visualization.
Data from 10 speakers of 11 genres are selected to generate speaker vectors.
The t-SNE toolkit~\cite{saaten2008} is applied to project the speaker vectors to a 2-dimensional space.
Figure~\ref{fig:ti} and Figure~\ref{fig:tx} present the distribution of i-vectors and x-vectors, respectively.

In Figure~\ref{fig:ti}, it can be seen that with i-vectors, speakers are largely intermingled with each other.
This is not surprising as the i-vector model is purely unsupervised and reflects variations of both speaker traits and acoustic conditions. 
Therefore, it is naturally hard to discriminate amongst speakers in multi-genre conditions.

For x-vectors presented in Figure~\ref{fig:tx}, one can observe larger inter-speaker distance and smaller intra-speaker distance compared to i-vectors.
This indicates that the x-vector model has its advantage to tackle the acoustic complexity associated with multiple genres.
Nevertheless, the genre complexity still leads to complicated intra-speaker distributions and overlap among different speakers.
This demonstrates that multi-genre speaker recognition is quite challenging.

\section{Experiment II: Multi-genre training}
\label{sec:exp2}

A straightforward approach to improve the performance under multi-genre conditions is to train
the speaker recognition system using multi-genre data, called \textbf{multi-genre (MG) training}.
Correspondingly, training using single-genre data (e.g., VoxCeleb) is called \textbf{single-genre (SG) training}.
In the following experiments, we use CN-Celeb.T to denote the speech data in CN-Celeb but not in CN-Celeb.E.
We use VoxCeleb for SG training, and CN-Celeb.T ($2,800$ speakers in total) for MG training.

Besides, as shown Figure~\ref{fig:multi},
there is a large proportion of multi-genre speakers in CN-Celeb (also in CN-Celeb.T).
These multi-genre speakers are the most important for MG training, as their data can inform the model what variations are caused by genres.
In order to investigate the contribution of the multi-genre speakers (which to some extent are truly multi-genre data),
we relabel CN-Celeb.T such that the data from the same speaker for different genres are treated as originating from different speakers. This relabeled dataset is denoted by CN-Celeb.T/SI, where SI means speaker isolation.
We call CN-Celeb.T/SI as \textbf{partial multi-genre data}, and the training on CN-Celeb.T/SI as \textbf{partial MG training}.

In this experiment, we compare the overall performance on CN-Celeb.E with different training schemes.
Since the baseline system consists of two components: the i-vector/x-vector front-end model and the PLDA back-end model,
we investigate the impact of MG training on the two components respectively. The results are shown in Table~\ref{tab:res-mgt-overall}.

\begin{table*}[htb!]
\caption{Overall EER (\%) results with single-genre, multi-genre and partial multi-genre training.}
\label{tab:res-mgt-overall}
\centering
\scalebox{1}{
\begin{tabular}{llllll}
\cmidrule(r){1-6}
\multirow{2}{*}{System} & \multicolumn{1}{l}{\multirow{2}{*}{Scheme}} & \multicolumn{1}{l}{\multirow{2}{*}{Front-end}} & \multicolumn{1}{l}{\multirow{2}{*}{Back-end}} & \multicolumn{2}{c}{CN-Celeb.E} \\
\cmidrule(r){5-6}
& \multicolumn{1}{l}{} & \multicolumn{1}{l}{} & \multicolumn{1}{l}{} & \multicolumn{1}{l}{Cosine} & \multicolumn{1}{l}{PLDA} \\
\cmidrule(r){1-6}
i-vector
&  (a) & VoxCeleb      & VoxCeleb          & 20.88     & 18.37    \\
&  (b) & VoxCeleb      & CN-Celeb.T        & 20.88     & 15.30    \\
&  (c) & VoxCeleb      & CN-Celeb.T/SI     & 20.88     & 16.31    \\
&  (d) & CN-Celeb.T    & CN-Celeb.T        & 19.29     & 14.01    \\
&  (e) & CN-Celeb.T/SI & CN-Celeb.T/SI     & 19.25     & 14.81    \\
\cmidrule(r){1-6}

x-vector
&  (a) & VoxCeleb      & VoxCeleb          & 20.13     & 16.59    \\
&  (b) & VoxCeleb      & CN-Celeb.T        & 20.13     & 13.44    \\
&  (c) & VoxCeleb      & CN-Celeb.T/SI     & 20.13     & 14.76    \\
&  (d) & CN-Celeb.T    & CN-Celeb.T        & 20.35     & 12.52    \\
&  (e) & CN-Celeb.T/SI & CN-Celeb.T/SI     & 20.83     & 13.65    \\
\cmidrule(r){1-6}
\end{tabular}}
\end{table*}

\subsection{Front-end model training}

For the front-end models, we firstly compare the performance between SG training (a) and MG training (d).
To eliminate the impact of the back-end model, we just look at the results with cosine scoring.
It can be observed that MG training does not give a clear advantage over SG training, especially with the x-vector model (20.35\% vs. 20.13\%).
This may be attributed to the bias in speaker numbers (2,800 in CN-Celeb.T vs. 7,000$+$ in VoxCeleb).
Note that with the i-vector model, the MG training leads to slightly better performance than the SG training (19.29\% vs. 20.88\%) although
the MG training used much less data. This better performance could be explained by the fact that
the training data (CN-Celeb.T) and the test data (CN-Celeb.E) are coherent in the MG training, in both languages and acoustic conditions.
This coherence is important for the i-vector model that is generative and descriptive.

Secondly, we compare the performance between MG training (d) and partial MG training (e).
We again focus on the cosine scoring.
Due to the unsupervised training strategy of the i-vector model, MG training and partial MG training obtain the same EER results.
For the x-vector model, partial MG training is a bit inferior to MG training (20.83\% vs. 20.35\%).
This is expected as the speaker labels of the partial multi-genre data lose the cross-genre information after
speaker isolation.

\subsection{Back-end model training}

To investigate the impact of MG training on the back-end PLDA model,  we compare the performance between SG training (a) and MG training (b).
It can be seen that performance with PLDA scoring improves with the MG training, for both the i-vector and x-vector systems.

Secondly, when comparing the training scheme (c) to (a) and (b), it can be seen that although the partial MG training is worse than the MG training
(14.76\% vs. 13.44\% for the x-vector system), it greatly outperforms the SG training (14.76\% vs. 16.59\% for the x-vector system).
This indicates that even without cross-genre speakers (true multi-genre data), data collected from multiple genres
are still very useful. This is good news, as collecting partial multi-genre data is much cheaper than collecting true multi-genre data.

In summary, the multi-genre data is important for MG training and can improve the performance on multi-genre test.
The MG training can be employed to either the front-end model or the back-end PLDA; though the best performance
is obtained when both are MG trained. Partial MG training is not as effective as the true MG training, but it can provide reasonable
gains with a low cost.

\section{Conclusion}

In this paper, we presented a comprehensive study for multi-genre speaker recognition.
To make the study feasible, we firstly collected and published a large-scale multi-genre corpus, CN-Celeb2.
Combined with the previously published CN-Celeb1, we have sufficient data to train and test
speaker recognition systems in multi-genre conditions.

Based on the new dataset, we firstly evaluated the performance of the state-of-art speaker recognition systems
in the multi-genre scenario, through which we identified the most difficult genres, and demonstrated that
the major challenge of multi-genre speaker recognition lies in both genre complexity and genre mismatch.
In the second experiment, we employed multi-genre training to tackle the multi-genre difficulties.
Significant performance improvement was obtained, and importance of multi-genre speakers was identified.


Multi-genre speaker recognition is very important but is also very challenging. The research presented in this paper should be regarded as
an initial and preliminary study in this direction. Lots of work remains to be done on this topic; to mention a few:  (1) collection of more multi-genre data to support the
research; (2) development of more powerful front-end models in order to produce genre-independent vectors; (3) discovery of more powerful back-end models to
handle the changed statistics from one genre to another; (4) exploration of physiological models that can describe the intrinsic change of human pronunciation in different genres.
We anticipate that the multi-genre challenge will be one of the prime
obstacles that needs to be tackled before the speaker recognition techniques find ubiquitous applicability in practice.


\bibliographystyle{IEEEtran}
\bibliography{cas-refs}

\end{document}